\documentclass[12pt]{article}
\usepackage{hyperref}
\usepackage{amsmath,amssymb,dsfont}
\usepackage{url,color}
\usepackage{graphicx}
\usepackage{subcaption}
\makeatletter
\newif\if@preliminary
\@preliminaryfalse
\def\preliminary{\@preliminarytrue}

\def\preprintno#1{\def\@preprintno{#1}}
\def\address#1{\def\@address{#1}}
\def\email#1#2{\thanks{\tt #1@{}#2}}
\def\abstract#1{\def\@abstract{#1}}
\renewcommand\abstractname{ABSTRACT}
\newlength\preprintnoskip
\newlength\abstractwidth
\@titlepagetrue
\renewcommand\maketitle{\begin{titlepage}%
  \let\footnotesize\small
  \hfill\parbox{\preprintnoskip}{%
  \begin{flushright}\@preprintno\end{flushright}}\hspace*{1cm}
  \vskip 60\p@
  \begin{center}%
    {\Large\bf\boldmath \@title \par}\vskip 1cm%
    {\sc\@author \par}\vskip 3mm%
    {\@address \par}%
    \if@preliminary
      \vskip 2cm {\large\sf PRELIMINARY DRAFT \par \@date}%
    \fi
  \end{center}\par
  \@thanks
  \vfill
  \begin{center}%
    \parbox{\abstractwidth}{\centerline{\abstractname}%
    \vskip 3mm%
    \@abstract}
  \end{center}
  \end{titlepage}%
  \setcounter{footnote}{0}%
  \let\thanks\relax\let\maketitle\relax
  \gdef\@thanks{}\gdef\@author{}\gdef\@address{}%
  \gdef\@title{}\gdef\@abstract{}\gdef\@preprintno{}
}%
\def\@citex[#1]#2{\if@filesw\immediate\write\@auxout{\string\citation{#2}}\fi
  \def\@citea{}\@cite{\@for\@citeb:=#2\do
    {\@citea\def\@citea{,\penalty\@m}\@ifundefined
       {b@\@citeb}{{\bf ?}\@warning
       {Citation `\@citeb' on page \thepage \space undefined}}%
\hbox{\csname b@\@citeb\endcsname}}}{#1}}
\def\citerange{\@ifnextchar [{\@tempswatrue\@citexr}{\@tempswafalse\@citexr[]}}
\def\@citexr[#1]#2{\if@filesw\immediate\write\@auxout{\string\citation{#2}}\fi
  \def\@citea{}\@cite{\@for\@citeb:=#2\do
    {\@citea\def\@citea{--\penalty\@m}\@ifundefined
       {b@\@citeb}{{\bf ?}\@warning
       {Citation `\@citeb' on page \thepage \space undefined}}%
\hbox{\csname b@\@citeb\endcsname}}}{#1}}
\long\def\@makecaption#1#2{%
  \vskip\abovecaptionskip
  \sbox\@tempboxa{#1: \emph{#2}}%
  \ifdim \wd\@tempboxa >\hsize
    #1: \emph{#2}\par
  \else
    \hbox to\hsize{\hfil\box\@tempboxa\hfil}%
  \fi
  \vskip\belowcaptionskip}
\makeatother

\newcommand{\ii}{\mathrm{i}}

\newcommand{\LL}{\mathcal{L}}

\newcommand{\vT}{\mathbf{T}}
\newcommand{\vW}{\mathbf{W}}
\newcommand{\vH}{\mathbf{H}}

\newcommand{\tr}[1]{\operatorname{tr}\left[#1\right]}
\newcommand{\vB}{\mathbf{B}}
\newcommand{\vD}{\mathbf{D}}

\newcommand{\GeV}{\text{GeV}}
\newcommand{\TeV}{\text{TeV}}
\newcommand{\ab}{\text{ab}}

\makeatletter
\font\manfnt=manfnt
\def\Watchout{\@ifnextchar [{\W@tchout}{\W@tchout[1]}}
\def\W@tchout[#1]{{\manfnt\@tempcnta#1\relax%
  \@whilenum\@tempcnta>\z@\do{%
    \char"7F\hskip 0.3em\advance\@tempcnta\m@ne}}}
\def\remark{\@ifnextchar[{\@remark}{\@remark[1]}}

\def\@remark[#1]#2{%
  \setbox\@tempboxa\hbox{\W@tchout[#1]}
  \@tempdima\wd\@tempboxa
  \list{}%
    {\leftmargin\@tempdima}%
    \item[\hbox to 0pt{\hss\W@tchout[#1]}]%
    \textbf{[#2]}}
\makeatother

\begin{document}


\preprintno{%
  DESY 16-098\\
  SI-HEP-2016-21 \\
  KA-TP-11-2016
}

\title{%
  Scattering of $W$ and $Z$ Bosons\\ at High-Energy Lepton Colliders
  }

\author{
  Christian Fleper\email{christian.fleper}{uni-siegen.de}$^a$,
  Wolfgang Kilian\email{kilian}{physik.uni-siegen.de}$^a$,
  J\"urgen Reuter\email{juergen.reuter}{desy.de}$^b$,
  Marco Sekulla\email{marco.sekulla}{kit.edu}$^c$
}

\address{\it%
    $^a$University of Siegen, Department of Physics,
    D--57068 Siegen, Germany, \\
    $^b$DESY Theory Group,
    D--22607 Hamburg, Germany,\\
    $^c$Institute for Theoretical Physics, Karlsruhe Institute of
    Technology, D--76128 Karlsruhe, Germany
}

\date{%
  \today
}

\abstract{%
  We present a new study of quasi-elastic $W$ and $Z$
  scattering processes in high-energy $e^+e^-$ collisions,
  based on and extrapolating
  the low-energy effective theory which extends the Standard Model
  with a $125\;\GeV$ Higgs boson.
  Besides parameterizing deviations
  in terms of the dimension-8
  operators that arise in the effective theory,
  we also study simplified models of new physics in $W/Z$
  scattering in terms of scalar and tensor resonance multiplets.
  The high-energy asymptotics of all
  models is regulated by a universal unitarization procedure.  This
  enables us to provide benchmark scenarios which can be
  meaningfully evaluated off-shell and in exclusive event samples, and
  to determine the sensitivity of an $e^+e^-$ collider to the model
  parameters. We analyze the longitudinal vector boson scattering
  modes, where we optimize the cuts for the fiducial cross section
  for different collider scenarios. Here, we choose energy stages of
  1.0, 1.4 and 3 TeV, as motivated by the extendability of the ILC
  project and the staging scenario of the CLIC project.
}
\maketitle


\tableofcontents


\section{Introduction}
\label{sec:intro}

Quasi-elastic scattering processes of
the massive electroweak bosons $W^\pm,Z$ (vector-boson scattering,
VBS) are a cornerstone in the
phenomenology of electroweak interactions.  The longitudinal
polarization components of on-shell energetic $W$ and $Z$ particles
are closely related to the unobservable Goldstone bosons that
constitute the elementary Higgs doublet together with the physical
Higgs boson~\cite{Chanowitz:1985hj,Gounaris:1986cr}.
Their interactions thus probe the mechanism of
electroweak symmetry breaking.

If the Higgs was missing from the
particle spectrum, the $W/Z$ interaction strength would rise with
energy into a non-perturbative regime, indicating an intrinsic cutoff
of the effective theory and new strong
interactions~\cite{Lee:1977yc,Lee:1977eg}.
However, the recent discovery of
a $125\;\GeV$ Higgs boson implies that this is not the case.  The
existence of the Higgs boson allows all VBS
interactions to remain weak asymptotically, calculable in
electroweak perturbation theory.  For the necessary cancellation of terms to
take effect to all orders, the vector-boson, Higgs and Goldstone
couplings must coincide exactly with their Standard Model (SM)
values~\cite{Cornwall:1974km}.

Without sufficient experimental data,
we cannot decide whether the pure SM is the correct description of the
electroweak symmetry-breaking sector.
Beyond the SM, processes that involve Higgs and Goldstone fields could
rather open a
new portal to phenomena and structures that do not couple
directly to matter particles, and are thus detached from immediate
experimental access.  New physics in the Higgs-Goldstone sector could
have only a
tiny impact on low-energy precision observables and is therefore largely
unconstrained by existing data.  Collider experiments will have to
search for new
effects in this area and complete our knowledge about the
particle and interaction spectrum at accessible energies.

VBS processes have been observed at the Large Hadron Collider (LHC) in Run
I~\cite{Aad:2014zda,ATLAS:2014rwa,CMS:2015jaa}, and the analysis
of future LHC runs at full energy and increased luminosity will
considerably improve our knowledge in this sector.  At hadron
colliders, VBS is accessible in processes of the type $pp\to q^\ast q^\ast\to
qq+VV$, where $q$ indicates any light quark, and $V=W,Z$.  The analysis
has been demonstrated to be feasible, but nevertheless
suffers from limitations that do not simply disappear with increasing
collider luminosity or energy.  Leptonic decays of vector bosons
either yield incomplete kinematic information ($W$) or a low branching
ratio ($Z$).  Hadronic decays are difficult to isolate from a large
QCD background.  VBS processes must be separated from
vector-boson pair production, $q\bar q'\to VV$, QCD production of $VV$
and two jets, as well as top-quark production as SM background.  The
initial state, i.e., the flavor and energy of the initial quarks,
cannot be controlled or detected, and steeply falling parton structure
functions limit the accessible energy for the elementary VBS
interaction.

By contrast, $e^+e^-$ colliders provide a clean environment where the
VBS process class has a unique signature, and most
decays of $W$ and $Z$ bosons are accessible and can be observed with high
efficiency and purity~\cite{Accomando:1997wt,Abe:2001swa}.
The initial state is known exactly, and the
$e^+e^-$ c.m.\ energy is fixed up to minor electromagnetic radiation
effects.  We expect more detailed and complementary information
to be available from $e^+e^-$ collisions compared to hadronic
collisions.  The decisive factor for an $e^+e^-$ collider is the
availability of sufficient energy combined with high luminosity.

The complete list of
$e^+e^-$ VBS processes accessible at an $e^+e^-$
collider with sufficient energy reads
\begin{align}
  e^+e^- &\to \bar\nu_e\nu_e W^+W^-,
\\
   &\quad \bar\nu_e\nu_e Z Z,
\\
   &\quad \bar\nu_e e^- W^+ Z,
\\
   &\quad e^+\nu_e Z W^-,
\\
   &\quad e^+e^- W^+W^-,
\\
   &\quad e^+e^- ZZ.
\end{align}
with various decay channels of the final-state bosons.
These processes allow us to study interactions of $W$ and $Z$
bosons in the following elementary scattering channels, which are realized as
distinct, approximately on-shell factorized contributions to the amplitudes:
\begin{align}
  W^+ W^- &\to W^+ W^-, \\
  W^+ W^- &\to ZZ, \\
  W^\pm Z & \to W^\pm Z, \\
  ZZ &\to W^+W^-, \\
  ZZ &\to ZZ.
\end{align}
By kinematics alone, the threshold for these processes is just above the
vector-boson pair production threshold of $160\ldots 180\,\GeV$, but for a
meaningful analysis of deviations from the SM, a significantly larger
vector-boson pair energy (i.e., invariant mass of the
$VV$ system) is necessary.

The prospects for the measurement of VBS processes at lepton
colliders, including potential non-SM contributions, have been studied
extensively in the
literature~\cite{Gunion:1987ta,TofighiNiaki:1988bz,Barger:1995cn,Han:1997ht,Dominici:1997zh,Boos:1997gw,Boos:1999kj,Chierici:2001ar,Rosati:2002wqa,Beyer:2006hx}.
However, most
previous studies have investigated no-Higgs (or heavy-Higgs) scenarios,
occasionally in comparison to the pure SM with a light Higgs of some
arbitrarily assumed mass.  Since a Higgs-like boson has been found,
its mass has been precisely determined, and the measurement of its
couplings is in accordance with the pure SM prediction, Higgs-less
models do no longer provide a viable scenario
for electroweak interactions.  The analysis of VBS should rather be
based on models that reduce to the SM with a $125\;\GeV$ Higgs at low
energy and take into account all recently accumulated knowledge about the
Higgs boson.\footnote{Turning this around, VBS at high energy can be
  utilized to supply indirect information on Higgs boson
  properties~\cite{Liebler:2015aka}.}

In this paper, we therefore present a new study of VBS
processes at an
$e^+e^-$ lepton collider.  We take the SM with a Higgs mass of
$125\;\GeV$ as reference and determine the sensitivity to effects beyond
the SM in VBS interactions.
Given the exploratory nature of this
task and the emergence of strong-interaction effects, we can restrict
our calculation to leading order in the SM and EFT
predictions.  Actual data analysis should incorporate NLO
corrections in the SM and its EFT extension,
cf.~\cite{Denner:1996ug,Denner:1997kq,Accomando:2006hq},
which eventually should become available for the complete six-fermion
partonic processes.  For a meaningful extrapolation to
energy and parameter ranges where partial-wave unitarity becomes an
issue, we apply a universal unitarization
method, the T-matrix framework.  The concrete
results are computed for specific values of the collider energy,
$\sqrt{s}=1.4\;\TeV$ and $3.0\;\TeV$ with integrated luminosities of
$\mathcal{L}_{int}=~1.5\;\ab^{-1}$ and $2\;\ab^{-1}$, as planned for
the Compact Linear Collider
(CLIC)~\cite{Linssen:2012hp,Lebrun:2012hj,Aicheler:2012bya}.
Furthermore, we make use of specific properties of the CLIC collider
environment in its currently planned state, including the most relevant
detector properties,
to estimate the sensitivity on anomalous effects beyond the
SM. For completeness, we also include numbers for a lower collider
energy of $1\;\TeV$ with an integrated luminosity of
$\mathcal{L}_{int}=~5\;\ab^{-1}$, which should also illuminate
the potential of an energy upgrade of the International Linear Collider
(ILC)~\cite{Baer:2013cma,Behnke:2013lya}.

\section{Effective Field Theory and Vector Boson Scattering}

The theoretical basis for the current study is the SM with a single
complex Higgs doublet.  Since we want to provide not just the unique
SM prediction but a range of possibilities for the high-energy
behavior of electroweak interactions, we have to
regard the SM as an effective field theory (EFT).
I.e., we assume an infinite series of interactions, organized by
operator dimension~\cite{Appelquist:1974tg,Weinberg:1978kz,Georgi:1994qn}.  The pure SM limit consists of setting the
couplings of all operators with dimension greater than four to zero,
or alternatively, letting the intrinsic mass scale of those
dimensionful couplings to infinity.

Within the EFT formalism, we assume that the strong-electroweak
$SU(3)_{QCD}\times SU(2)_L\times U(1)_Y$ gauge symmetry of the SM
Lagrangian is a fundamental property, and therefore organize all
higher-dimensional operators in terms of gauge-invariant polynomials.
This implies global strong-electroweak invariance for all terms and
promoting partial derivatives to covariant derivatives where required
by the field representation.  We do not consider CP
symmetry breaking effects (beyond those already present in the SM) in
the current study.

The SM Higgs, in the gauge-invariant Lagrangian, appears as a complex
doublet which transforms linearly under $SU(2)_L\times
U(1)_Y$.\footnote{A non-linear Higgs representation might be chosen to allow
  for more freedom in Higgs couplings that are not yet
  constrained with precision, but this property is
  of secondary importance in our study of VBS processes.  It becomes
  more relevant when considering also
  Higgs final states, cf.~\cite{Contino:2013gna}.}
If we neglect
the interactions of heavy SM fermions, the global symmetry is
approximately $SU(2)_L\times
SU(2)_R$~\cite{Veltman:1977kh,Sikivie:1980hm}.  This symmetry is
explicitly
broken by the hypercharge gauge coupling.  The obstruction can be
controlled as a spurion, and the analysis can be based on an exactly
$SU(2)_L\times SU(2)_R$ symmetric Lagrangian.  This is a reasonable simplification since
we are
interested in the high-energy range of VBS processes where gauge
interactions play a minor role, i.e., can be viewed as a perturbation
entirely.  The Higgs vacuum expectation value (vev) spontaneously
breaks the global symmetry down to the diagonal $SU(2)_{L+R}$, known
as the custodial symmetry $SU(2)_C$.

In this context, it is natural to represent the Higgs field multiplet
as a $2\times 2$ matrix~\cite{Kilian:2003pc,Kilian:2014zja},
\begin{align}
	  &&\vH &=
  \frac 1 2
  \begin{pmatrix}
    v+ h -\ii w^3 & -\ii \sqrt{2} w^+ \\
		-\ii \sqrt{2} w^- & v + h + \ii w^3  \\
  \end{pmatrix} \quad .
\end{align}
which transforms linearly under $U\in SU(2)_L$ and $V\in
SU(2)_R$ as
\begin{equation}
  \vH \to U\vH V^\dagger \quad .
\end{equation}
The covariant derivative of the Higgs matrix is given by
\begin{equation}
  \vD_\mu \vH = \partial_\mu \vH - i g \vW_\mu \vH + i g'\vH\vB_\mu
\end{equation}
where
\begin{equation}
  \vW_\mu \equiv W_\mu^a\frac{\tau^a}{2},
\qquad
  \vB_\mu \equiv B_\mu\frac{\vT}{2}
\end{equation}
and $\vT$ is a $SU(2)_R$-breaking spurion with ground state
$\vT=\tau^3$.  Finally, we define the matrix-valued field
strengths
\begin{equation}
    \vW^{\mu\nu} =\partial_\mu \vW_\nu - \partial_\nu \vW_\mu
    - \ii g \left [ \vW_\mu , \vW_\nu \right ],
\quad
    \vB^{\mu\nu} =\partial_\mu \vB_\nu - \partial_\nu \vB_\mu
\end{equation}
and quote the bosonic part of the SM Lagrangian (omitting QCD),
\begin{align}
  \begin{aligned}
    \LL_{SM} =
    &-\frac{1}{2}\tr{\vW_{\mu\nu}\vW^{\mu\nu}}
    -\frac{1}{2}\tr{\vB_{\mu\nu}\vB^{\mu\nu}} \\
    &+\tr{ \left ( \vD_\mu \vH \right )^\dagger\vD^\mu \vH }
    + \mu^2\tr{\vH^\dagger \vH}
    -\frac{\lambda}{2}\left( \tr{\vH^\dagger \vH} \right)^2
  \end{aligned} \quad .
  \label{SM-L}
\end{align}

The power of an EFT series as a perturbative expansion lies in the
accuracy of a truncation at low order.
New mutual and self-interactions of bosons only take the form of
gauge-invariant operators of even dimension.  The leading non-SM order
is thus dimension six, so it appears natural to truncate the EFT at this order.
The complete set of dimension-six
operators has
been discussed extensively in the
literature~\cite{Buchmuller:1985jz,Hagiwara:1992eh,Hagiwara:1993ck,Eboli:2006wa,Grzadkowski:2010es,Baak:2013fwa}.

However, new-physics effects in the
Higgs-Goldstone sector, including new strong interactions, largely decouple
from precision observables as they are accessible today.
While several dimension-six operators in the EFT do affect VBS processes at tree level,
they simultaneously
modify other couplings which can be measured independently.  Such operators are
an inadequate representation of the specific new-physics scenarios that VBS is
most sensitive to.
In fact, for the purpose of
studying VBS, we may
assume that the coefficients of
dimension-six operators are known to
sufficient precision; we set them to zero as the standard reference
point.

Traces of genuine new Higgs-Goldstone physics
appear first in dimension-eight operators.  These modify VBS interactions
independently of other types of interaction.  For instance, the
exchange of massive resonances generically results in dimension-eight
effective operators.  Our study therefore
includes the relevant dimension-eight operators.

It is evident that any truncation at this level, while technically
consistent, is of questionable value.  Amplitudes modified by
dimension-eight operators rise rapidly with energy and thus
yield large effects~\cite{Kilian:2014zja}.  This fact also indicates the breakdown of the
perturbative expansion.
Our version of the EFT with
dimension-eight operators is therefore not intended as an
approximation that can be systematically improved.  It rather serves as a
phenomenological tool
to describe dominant non-SM phenomena in VBS.  At low energy,
$O(100\;\GeV)$, it smoothly matches to the EFT within its range of validity.
Extrapolated to high energy, it gives a flavor of the
maximum impact that BSM can have on VBS.  The extrapolation
incorporates unitarity as the only
damping mechanism and thus asymptotically approaches the unitarity
bound for each partial-wave amplitude.  However, additional structure can easily be
included, as we will describe below using simplified models containing
resonances.

\section{Anomalous Interactions of Vector Bosons}

The complete set of dimension-eight operators for the SM fields is
rather large.  In line with the above considerations, we resort to standard
simplifications in order to make the set manageable.  First of all, as
stated before, we only consider bosonic operators.  (That is, we treat
fermionic currents as external probes for bosonic interactions,
neglecting genuine fermionic anomalous contributions.)  We furthermore
assume the global $SU(2)_L\times SU(2)_R$ symmetry of the fermion- and
gaugeless SM to hold also at higher orders.  This introduces relations
between $W$ and $Z$ scattering amplitudes and simplifies the
scattering matrix for the purpose of unitary extrapolation.

Focusing on operators that directly modify quartic vector-boson
interactions, there are three classes which affect Higgs/Goldstone
bosons only (index: $S$), gauge bosons only (index: $T$), or both
(index: $M$), respectively.  We recall that, taking EWSB into account, Goldstone
bosons are probed via the longitudinal polarization direction of
energetic $W$ and $Z$ bosons, while gauge bosons translate into
transversal polarization.

Operators involving only the Higgs/Goldstone bosons sector, i.e., S-type, are
represented by a combination of covariant derivatives acting on the
Higgs fields.    Two linearly independent
operators
can be identified
\begin{subequations}
  \label{eq:dim8_operator_S}
  \begin{alignat}{4}
    \LL_{S,0}&=
    & &F_{S,0}\ &&
    \tr{ \left ( \vD_\mu \vH \right )^\dagger \vD_\nu \vH}
    \tr{ \left ( \vD^\mu \vH \right )^\dagger \vD^\nu \vH},
    \\
    \LL_{S,1}&=
    & &F_{S,1}\ &&
    \tr{ \left ( \vD_\mu \vH \right )^\dagger \vD^\mu \vH}
    \tr{ \left ( \vD_\nu \vH \right )^\dagger \vD^\nu \vH}.
  \end{alignat}
\end{subequations}

In the present paper, we do not study the transversal and mixed
interactions but defer this to
future work.  For completeness, we list the $T$ and $M$-type operators
in the Appendix.  We also neglect operators which are proportional to
$\vH^\dagger \vH $; as long as Higgs final states are not considered,
this class of operators merely renormalizes the dimension-six part of
the EFT Lagrangian.

We may compare this EFT framework with the corresponding expansion~\cite{Longhitano:1980tm,Appelquist:1993ka} and
subsequent analysis for the no-Higgs (or heavy-Higgs) case, as it was
investigated, for instance, in~\cite{Boos:1997gw,Boos:1999kj}.  In the pure SM, a light
Higgs boson and no higher-order operators, the amplitude for VBS remains
perturbative at all energies.  Beyond the SM, new physics that appears exclusively in
the Higgs-Goldstone sector corresponds to dimension-eight operators as argued above,
therefore any deviation from the SM increases rapidly with energy.
Since the pure SM
amplitude is small, interference plays a minor role and the
phenomenologically relevant behavior originates from the
dimension-eight interactions, squared.  By contrast, if the light
Higgs boson did not exist, the reference amplitude would grow with energy and render the
interaction strong in the $\TeV$ range by itself, while anomalous
effects
would contribute first via their interference, which is a less
striking modification of the reference amplitude.  In this study, which
is based on the now established light-Higgs scenario, we
therefore will face a sudden transition from a
weakly-interacting to a strongly-interacting regime when increasing
the energy in the VBS process.

\section{Unitary Extrapolation}

The SM amplitudes for VBS, as for any other elementary process, are
perturbative throughout the accessible energy range and therefore do
not pose a unitarity problem.  If anomalous effects are present which
cannot simply be mapped to an extended renormalizable (weakly
interacting) model, this property is lost, and we have to deal with
unitarity violation in the tree-level prediction.  Clearly, this
indicates the breakdown of perturbation theory.  On the other hand,
the quantitative analysis of actual data relies on the availability of
a prediction which depends on a suitable set of new parameters.  Such
a reference allows us to quantitatively study or exclude a
deviation from the SM.  In order to discard grossly unphysical
parameterizations, we have to ensure that any such extrapolation is at
least in accordance with unitarity, all assumed symmetries, and
smoothly matches to the low-energy effective theory

In~\cite{Kilian:2014zja,Sekulla:2015}, based on earlier work
in~\cite{Alboteanu:2008my}, we have described the T-matrix
unitarization  framework as a generic scheme that allows for a unitary
extrapolation of any model without introducing arbitrary artefacts in the
asymptotic regime.  We adopt this framework for the current paper.

At the fundamental level of the complete scattering matrix
$\mathbf{S}=1+\ii \mathbf{T}$ (T matrix),
the unitarization framework relies on the formula
\begin{align}
  \label{eq:tmatrix}
  \mathbf{T}(\mathbf{T}_0) &= \frac{1}{\mathrm{Re} \left (
      {\mathbf{T}_0}^{-1} \right ) -
    \frac{\ii}{2} \mathds{1}} \, ,
\end{align}
which transforms an arbitrary model of the scattering matrix
$\mathbf{T}_0$ into a unitary model of the scattering matrix
$\mathbf{T}$.  In particular, the set of amplitudes is invariant under
the transformation if it already respects unitarity.  For the
application of the formula, it is advantageous to
diagonalize the scattering matrix first and apply unitarization to
eigenamplitudes, since this amounts to a simple multiplication or
subtraction.  In the present context, this can be done most easily in
the high-energy limit where the gauge sector decouples from the Higgs
sector.  Since unitarity is an issue only for high energies, such an
approach is sufficient to remove all dangerous terms.

The method does not refer to any property of the original model which
is encoded in $\mathbf{T}_0$; in particular, it does not assume a
perturbative expansion or a particular analytical structure.  While it
reconciles any model $\mathbf{T}_0$ with unitarity, the result is not
a unique prediction.  It is merely a model that describes possible
behavior of the amplitude up to the highest energies, as opposed to a
unitarity-violating model or extrapolation that describes impossible behavior.

In practice, we identify the asymptotically leading terms in the
diagonal scattering matrix, replace them by their unitary equivalents,
and invert the diagonalization.  Factoring out Lorentz tensors and
subtracting the original EFT contribution, we cast this in the form of
extra momentum-dependent Feynman rules.  The new terms resemble the
local Feynman rules that describe the original EFT operators, but
they carry prefectors that depend on invariant momentum combinations
in a non-analytical form.  Nevertheless, for calculational purposes
the new terms play a role analogous to the local counterterms that
arise in a NLO calculation, and are straightforward to take into
account in the construction and evaluation of scattering matrix
elements.  This allows us to perform off-shell calculations and
evaluate the unitarized amplitudes for external fermions, as long as
the kinematic assumptions underlying the procedure are satisfied.

For
illustration, in plots below (Figs.~\ref{i:Pt}, \ref{i:Minv},
\ref{i:Rec})
we display both the unitarized
model behavior and the unphysical
results that we would have obtained without unitarization.  The latter are
marked as dashed lines, while continous curves refer to the
respective unitary versions.  While for some parameter sets, the
unitarization correction remains a minor issue, there are various
cases where unitarization has a large impact.  This property clearly
indicates that for CLIC energies, VBS processes are probed in a range
where undetermined new-physics effects are actually important, and
systematic approximations do no longer yield unambiguous results.
Under no circumstances, calculation results without unitarization may
be used for the analysis of actual data, as this would grossly
overestimate the sensitivity on the model parameters.

\section{Cut-Based Extraction of the VBS Signal}

Based on the theoretical framework as described above, we now turn to
the actual prospects for measuring vector-boson scattering
processes at high effective energy at a high-energy lepton collider,
like e.g. the CLIC collider.  The strategy
of such an analysis does not depend strongly on the underlying physics
model.  We can therefore follow the ideas of~\cite{Barger:1995cn,Boos:1997gw,Boos:1999kj} and
adapt the analysis to the CLIC environment.

\begin{figure}[h!]
 \centering
  \includegraphics[width=0.8\textwidth]{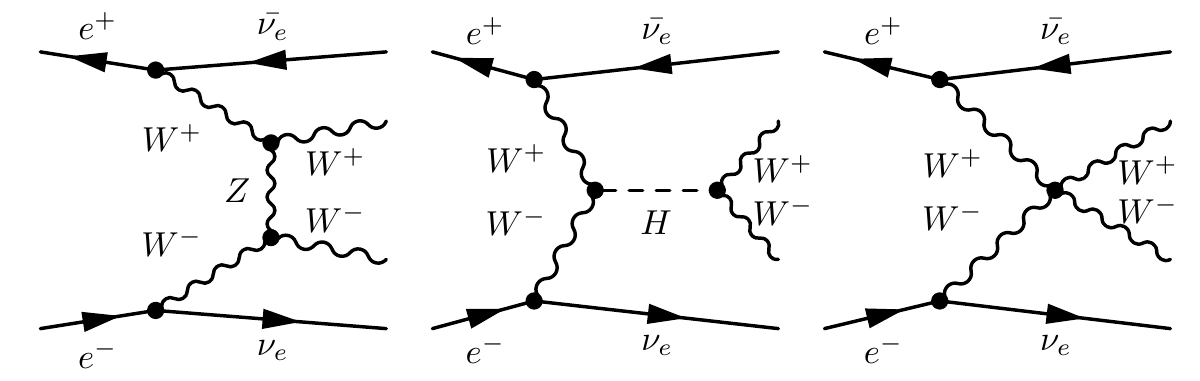}
 \caption{Signal Feynman diagrams contributing to the VBS process.}
 \label{i:signal}
\end{figure}

\begin{figure}[h!]
 \centering
  \includegraphics[width=0.8\textwidth]{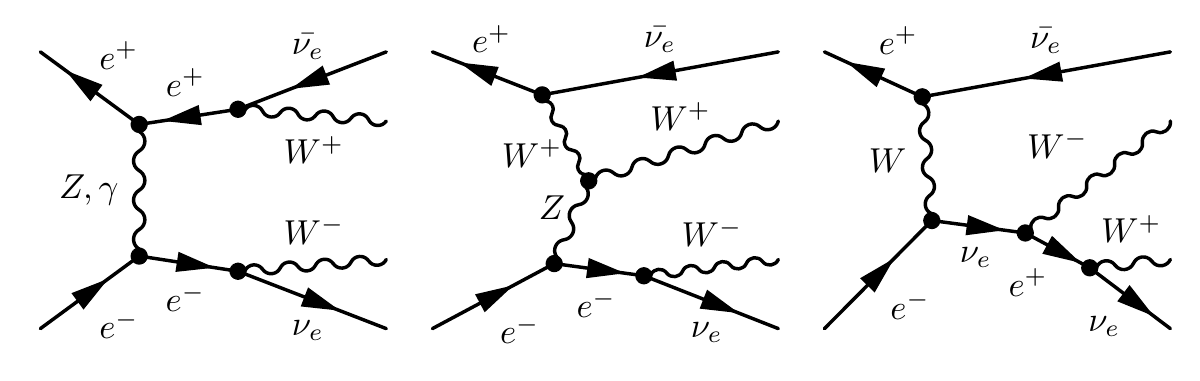}
 \caption{Feynman diagrams contributing to the irreducible background.}
 \label{i:irrbkg}
\end{figure}

\begin{figure}[h!]
 \centering
  \includegraphics[width=0.8\textwidth]{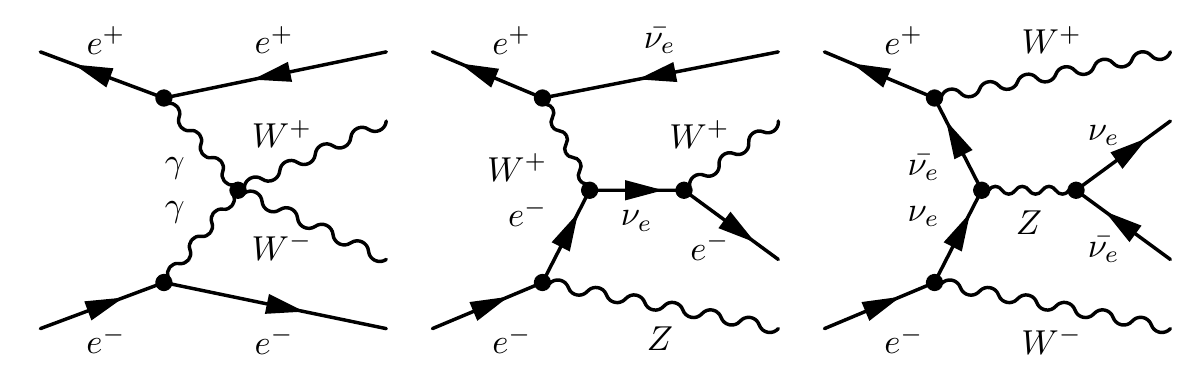}
 \caption{Feynman diagrams contributing to the partially reducible background.}
 \label{i:rbkg}
\end{figure}

The main idea is to isolate a signal of vector-boson pairs in
association with very forward neutrinos or electrons.  This is the
kinematic situation where in the diagrams of Fig.~\ref{i:signal},
the incoming vector bosons which participate in the VBS interaction
have small invariant mass, so the amplitude is enhanced by the small
denominator of the t-channel propagators.  The nonvanishing mass of
$V=W,Z$ bosons cuts off the approach to the actual particle pole, so
the enhancement factor is of order $m_V^2/E_V^2$ in this kinematic
range, and the typical recoil $p_T$ of the radiating lepton becomes of
order $m_V$.  Graphs without the VBS interaction
(Figs.~\ref{i:irrbkg}, \ref{i:rbkg}) do not benefit from this effect.

A large part of the total rate is due to $W$-boson pair production
initiated by photons, cf.~Fig.~\ref{i:rbkg} (left).  Photon-induced
VBS processes are interesting by themselves, but photons do not have a
longitudinal component and therefore do not receive contributions from
new physics in the Higgs sector.  For our purposes, the photon-induced
contribution is therefore considered as a background.  In contrast to
the distribution of intermediate $W/Z$ bosons, which is dominated by
$Q\sim m_{W/Z}$, the distribution of photons extends down to nearly
zero momentum transfer and is cut off only by the finite electron mass.
Electrons
in forward direction disappear in the beampipe and are thus indistinguishable
from neutrinos, so in this kinematic range, intermediate photons cannot be
separated by detecting the radiating fermion.  Nevertheless, this background
can be reduced without losing too much of the signal by
vetoing against very forward electrons and simultaneously cutting on the
$p_T$ distribution
of the vector-boson pair system.

We are interested in
the vector-boson pair system at large combined invariant mass, as this
is the energy value that enters the basic VBS interaction. Hence, we
propose the following set of selection cuts, where the first number
always refers to the 3.0 TeV CLIC staging, while the one in
parentheses corresponds to the lower-energy staging at 1.4 TeV.

\begin{enumerate}
 \item $M_{inv}(\bar{\nu}\nu)>230(175)\;\GeV$. The signal process
       contains two neutrinos in the final state. This cut removes
       events where the neutrinos originate from Z decay; a majority
       of these neutrinos will have an invariant mass
       $M_{inv}(\bar{\nu}\nu) \sim 91\;\GeV$.  Backgrounds
       from $W^+ W^-$ and QCD four-jet production are also removed by
       this cut.
 \item $|\cos \theta (W/Z)| < 0.8$ and $p_\perp (W/Z)>300(180)\;\GeV$.
       Events where the vector bosons have a small distance from the
       beam pipe or a small transverse momentum are cut away. This
       reduces backgrounds which result from t-channel exchange in the
       subprocess.
 \item $\theta (e)>15$ mrad and $p_\perp (WW)>100(50)\;\GeV$,
       $p_\perp (ZZ)>60(40)\;\GeV$. Together with a
       cut on the transverse momentum of the vector boson pair system,
       background resulting from $\gamma \gamma$ fusion
       will substantially be decreased.
 \item $900(800)\;\GeV < M_{inv} (WW) < 1900(1175)\;\GeV$,
       $850(800)\;\GeV < M_{inv} (ZZ) < 1900 (1175)$ GeV. The influence
       of the operators Eq.~(\ref{eq:dim8_operator_S}) increases with
       the invariant mass of the vector boson pair. We therefore only
       consider events within a small invariant mass range.
\end{enumerate}

The following cuts are applied for the ILC staging scenario at $1.0\,\TeV$:

\begin{enumerate}
\item $M_{inv}(\bar{\nu}\nu)>150\;\GeV$.
\item $|\cos \theta (W/Z)| < 0.8$ and $p_\perp (W/Z)>150\;\GeV$.
\item $\theta (e)>15$ mrad and $p_\perp (WW)>45\;\GeV$,
       $p_\perp (ZZ)>40\;\GeV$.
\item $575\;\GeV < M_{inv} (WW) < 800\;\GeV$,
       $600\;\GeV < M_{inv} (ZZ) < 800$ GeV.
\end{enumerate}

For the calculation, we have used the implementation of the SM
in the WHIZARD Monte-Carlo
generator~\cite{Kilian:2007gr,Moretti:2001zz}.
The SM extensions that we consider in later sections are likewise handled by
dedicated WHIZARD model implementations, taking into account unitarization
corrections where necessary.
The WHIZARD generator provides a physics simulation framework
that can include beam properties (polarization, beamstrahlung and
ISR), the complete partonic final state to leading and next-to-leading
order QCD, parton shower, and
hadronization, eventually complemented by detector-level tools and
analysis
methods~\cite{Kilian:2011ka,Christensen:2010wz,Kilian:2012pz,WHIZARD_NLO}.

For this exploratory study, we have
largely restricted ourselves to on-shell $W/Z$ bosons in the final
state and concentrate on total cross section values after cuts.
Polarization, ISR, and beamstrahlung are not taken into account for
the tables below.  Further below, we comment on possible improvements
if polarization and details of the partonic final state are taken into
account, using off-shell simulation for the latter.  We expect that a
future experimental study which can be performed using the WHIZARD
generator in connection with detector simulation, jet
algorithms, etc., should give more accurate results which we
nevertheless expect to lie in the numerical range as our final results
suggest. Clearly, especially ISR and beamstrahlung will lead to a
distortion of distributions and to a depletion of events in the
high-energy region, but we do not expect our results to fundamentally
change. Our first estimates which should be backed up by future
experimental studies confirmed that, and especially the number that
roughly half of the events is in the highest ten-percent energy bin
from Tab. 2.2 in~\cite{Linssen:2012hp} when beamstrahlung is actually
simulated.

The tables~\ref{t:totcswocuts} and \ref{t:totcswcuts} contain our
results for the SM cross sections of the processes of interest, before
and after cuts, respectively.  As argued above, the results apply to
on-shell bosons in the final state and do not refer to
detailed properties of the CLIC collider, with two notable exceptions
that we discuss in the following.

\begin{table}[h]
 \centering
  \begin{tabular}{l c c c}
  \hline
  Process & 1.4 TeV & 3 TeV & Factor\\
  \hline
  $W^+W^-\nu\bar{\nu}$ & 47.1 & 132 & 1 \\
  $W^+W^-e^+ e^-$ & 1570 & 3820 & 1\\
  $W^\pm Z e^\mp \nu$ & 138 & 408 & 0.136\\
  $ZZe^+ e^-$ & 3.78 & 4.70 & 0.019\\
  $W^+ W^- (Z \to \nu\bar{\nu})$ & 11.7 & 9.35 & 1 \\
  \hline
  $ZZ\nu\bar{\nu}$ & 15.7 & 57.5 & 1\\
  $ZZe^+ e^-$ & 3.78 & 4.70 & 1\\
  $W^\pm Z e^\mp \nu$ & 138 & 408 & 0.136\\
  $W^+W^-e^+ e^-$ & 1570 & 3820 & 0.019\\
  $ZZ (Z \to \nu\bar{\nu})$ & 0.484 & 0.237 & 1\\
  \hline
 \end{tabular}
 \caption{Standard model total cross sections in fb ($\pm 1 \%$ error) without
 cuts for center-of-mass energies of $\sqrt{s}=1.4\;\TeV$ and $3\;\TeV$.
 Both particle beams are unpolarized. Detection efficiencies and branching
 ratios are not included. All cross sections have to be multiplied by the
 factors in the fourth column to take the misidentification of vector bosons
 into account.}
 \label{t:totcswocuts}
\end{table}

\begin{table}[h]
 \centering
  \begin{tabular}{l c c c}
  \hline
  Process & 1.4 TeV & 3 TeV & Factor\\
  \hline
  $W^+W^-\nu\bar{\nu}$ & 0.119 & 0.790 & 1 \\
  $W^+W^-e^+ e^-$ & 0.000 & 0.000 & 1\\
  $W^\pm Z e^\mp \nu$ & 0.269 & 1.200 & 0.136\\
  $ZZe^+ e^-$ & 0.000 & 0.000 & 0.019\\
  $W^+ W^- (Z \to \nu\bar{\nu})$ & 0.039 & 0.610 & 1 \\
  \hline
  $ZZ\nu\bar{\nu}$ & 0.084 & 0.790 & 1\\
  $ZZe^+ e^-$ & 0.000 & 0.000 & 1\\
  $W^\pm Z e^\mp \nu$ & 0.288 & 1.590 & 0.136\\
  $W^+W^-e^+ e^-$ & 0.000 & 0.000 & 0.019\\
  $ZZ (Z \to \nu\bar{\nu})$ & 0.000 & 0.000 & 1\\
  \hline
 \end{tabular}
 \caption{Same as Tab.~\ref{t:totcswocuts}, but with cuts.}
 \label{t:totcswcuts}
\end{table}

\begin{enumerate}
\item
For suppressing the contribution from photon-induced processes, the
analysis has to rely on the ability to detect energetic electrons in
the very forward region, which are barely deflected by the photon
emission.  If such electrons are vetoed against, the ambiguity between
$W$ and $Z/\gamma$ bosons in the initial state is greatly reduced.
For the current study we have assumed that a veto against electrons is
possible down to an angle of 15 mrad~\cite{Idzik:2015oja}.
\item
The clean environment and triggerless operation of a lepton collider
allows us to detect
final-state vector bosons by their decay products in essentially all
channels.  For the current study, we have concentrated on the hadronic
decay channels.  These decays provide the major part of the decay
branching fractions and yield complete kinematic information.  In
particular, the momenta of the forward neutrinos can be inferred from
the missing energy and momentum.  A disadvantage of hadronic channels
is the absence of charge information and the finite jet-pair invariant
mass resolution, which adds on the natural decay width of $W$ and $Z$
bosons.  As a result, there is a probability for misidentification
between $W$ and $Z$.  We take this into account by the matrix for
identification probabilities as it was determined
in~\cite{Marshall:2012ry}:

\begin{subequations}
  \label{eq:WZid}
  \begin{align}
    W &\to 88 \% \ W, \ 12 \% \ Z \\
    Z &\to 12 \% \ W, \ 88 \% \ Z
   \end{align}
\end{subequations}

Partonic $WW$, $WZ$ and $ZZ$ final states therefore will be identified as
a $WW(ZZ)$ event with probabilities 77.4\%, 10.6\%, 1.4\%, which yields
the weighting factors 1:0.136:0.019 given in the final column of both
tables~\ref{t:totcswocuts}
and \ref{t:totcswcuts}. Taking the hadronic $W$ and $Z$ boson branching
ratios of 67.70\% and 69.91\% into account and including the di-lepton modes
of the Z boson (BR=6.729\% for $e^+e^-$ and $\mu^+ \mu^-$), the efficiencies
for detecting a $WW$, $WZ$, $ZZ$ pair originating from a partonic
$WW$, $WZ$, $ZZ$ final state are 35.4\%, 40.1\% and 45.5\%,
respectively.
\end{enumerate}

A significant part of the $VV\nu\nu$ final state is contributed by
triple vector boson production, where the third boson is a $Z$ with
invisible decay to neutrinos, cf.\ Fig.~\ref{i:rbkg} (right).
This might be considered as a separate background, and it is tempting
to calculate this via a separate calculation of triple vector boson
production with $Z$ decay.  However, the process is probed in a
kinematical region where the would-be $Z$ boson is far off shell.  The
distribution in this region cannot be defined in a
gauge-invariant way if we select only diagrams with a virtual $Z$
boson.   We have compared the results
of~\cite{Boos:1997gw,Boos:1999kj}, which were obtained for this
selection of Feynman diagrams
using
'tHooft-Feynman gauge for the gauge bosons,
with an analogous WHIZARD calculation where all amplitudes are
computed in unitarity gauge.  As one may expect, the two results
differ by a large factor in the signal region of high $\bar\nu\nu$
invariant masses, while coinciding on the $Z$ mass peak.  The momentum
factors in the unitarity-gauge propagators produce a gauge-dependent
excess which is cancelled against a matching piece if the full
gauge-invariant set of diagrams is taken into account.  In tables~\ref{t:totcswocuts} and \ref{t:totcswcuts}, we quote the unitarity-gauge results for the three-boson subprocess for
completeness, but we emphasize that after cuts, this number is unphysical
and actually irrelevant for the signal sensitivity; as a background,
it can be ignored
for all practical purposes.

In other words,
such a selection of Feynman graphs should not be attempted and a
triple-boson background contribution cannot be defined.  For the
remainder of our study
we only use results for the complete process which are gauge invariant
by construction.

Besides gauge-invariance issues,
we should note that the $VVZ$ contribution to the processes of
interest does not depend significantly on the coefficients of the
operators in Eq.~\ref{eq:dim8_operator_S}.  This is easily understood
if we remember that in the high-energy limit gauge and Goldstone
bosons decouple.  An intermediate off-shell $Z$ or $\gamma$ couples to
a massless fermion current and therefore behaves as a gauge boson.  A
Goldstone boson contribution is excited only via mixing, suppressed by
$m_Z/\sqrt{s}$. Indeed we observe that in the low-mass region for the
neutrino system where the $Z$ decay contributes, the dependence on the
anomalous parameters is negligible.  We can therefore cut on high $VV$
invariant mass without losing sensitivity and thus concentrate on the
genuine VBS topology.  This property also removes any need for
unitarizing the triple-production channel, since the SM contribution
is guaranteed to respect perturbative unitarity.

\section{Results for EFT Operator Coefficients}

We now consider the extrapolated EFT as the most straightforward
extension of the SM for VBS at high energy.  We use the SM
Lagrangian~(\ref{SM-L}) and add the two dimension-eight
operators~(\ref{eq:dim8_operator_S}) with their coefficients $F_{S,0}$
and $F_{S,1}$ as free parameters.  The calculation is analogous to the LHC
case~\cite{Kilian:2015opv}; for the numerical results, we use the model
implementation in the WHIZARD event generator and the automatic
calculation of unitarized tree-level amplitudes, distributions, and
cross sections.

At low energy, the free parameters $F_{S,0}$ and
$F_{S,1}$ are identified as expansion parameters in the EFT, within
the range of validity of the latter.\footnote{In
  the ATLAS analysis of VBS at the LHC~\cite{Aad:2014zda}, the notation
  $(\alpha_4,\alpha_5)$ was adopted instead, borrowed from the no-Higgs
  EFT~\cite{Appelquist:1993ka}, although the theory model included the
  light Higgs and unitarization, i.e., worked with the extrapolated
  EFT model of the present paper. For the relation of coefficients,
  cf.~\cite{Baak:2013fwa}.}  Where we extrapolate the EFT
to high energies beyond its validity range, we apply T-matrix
unitarization to the
calculated amplitude.  The unitarization correction will set in where the
extrapolated amplitude becomes strong.  In effect, we obtain a smooth
transition to high energy asymptotics with saturation of unitarity in all
partial waves.  This is a two-parameter model for a structureless
strongly interacting continuum.

\begin{figure}[p]
   \begin{subfigure}[t]{0.46\textwidth}
      \includegraphics[width=\textwidth]{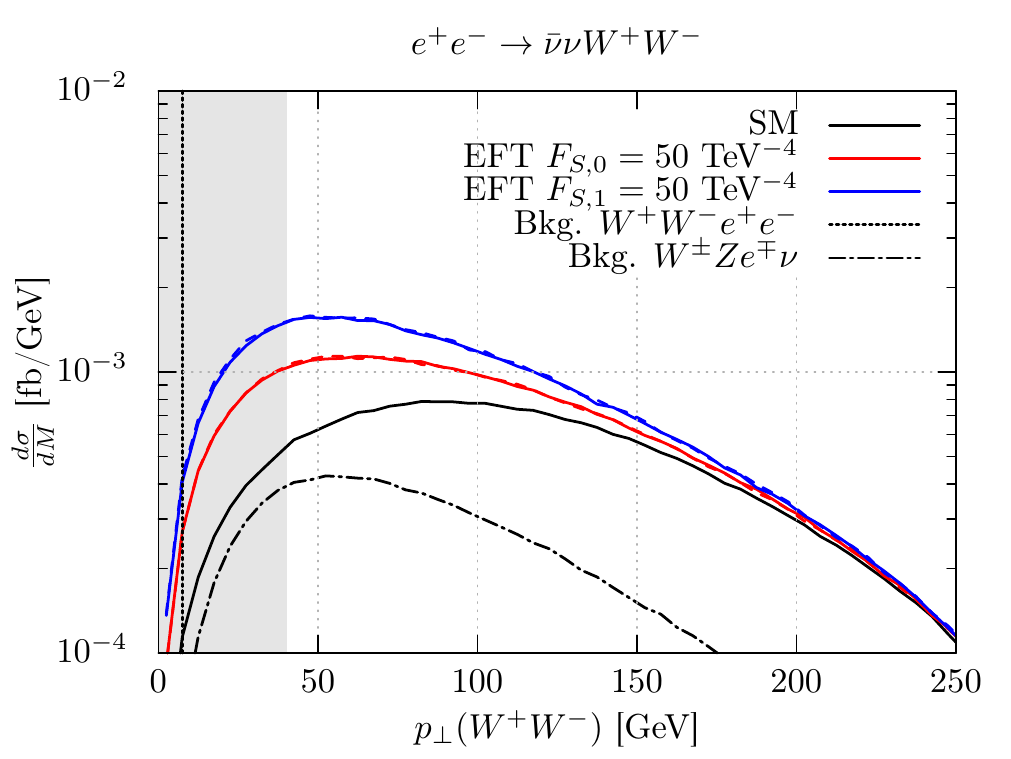}
   \end{subfigure}\hfill%
   \begin{subfigure}[t]{0.46\textwidth}
      \includegraphics[width=\textwidth]{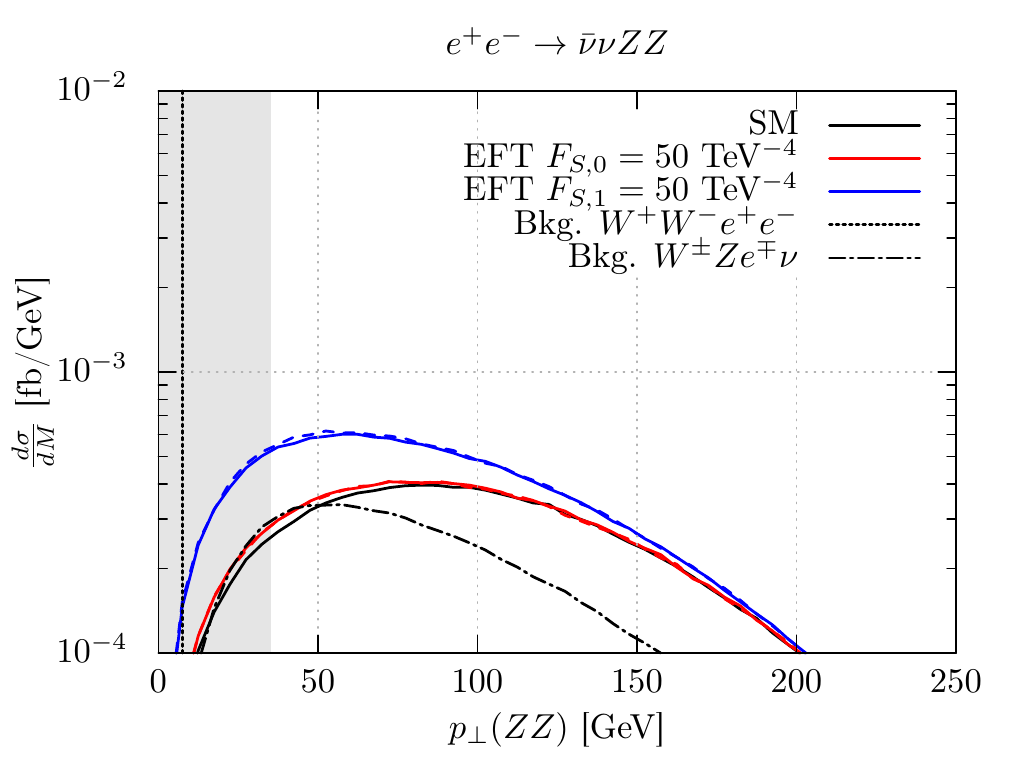}
   \end{subfigure}\\[1pt]%
   \begin{subfigure}[t]{0.46\textwidth}
      \includegraphics[width=\textwidth]{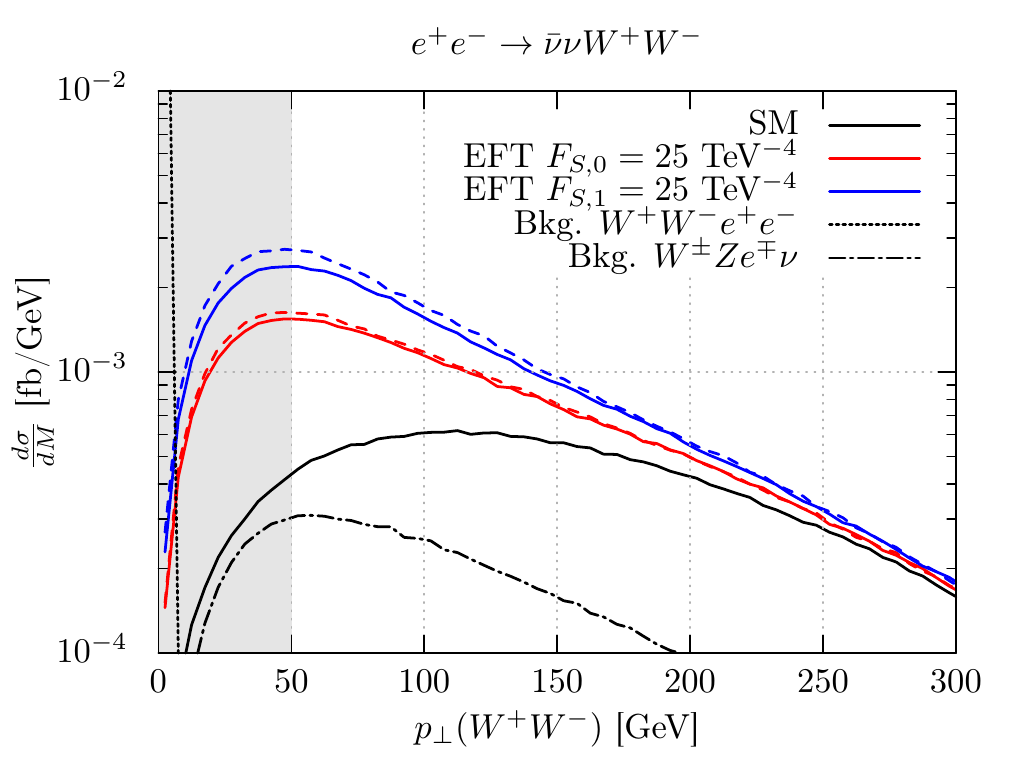}
   \end{subfigure}\hfill%
   \begin{subfigure}[t]{0.46\textwidth}
      \includegraphics[width=\textwidth]{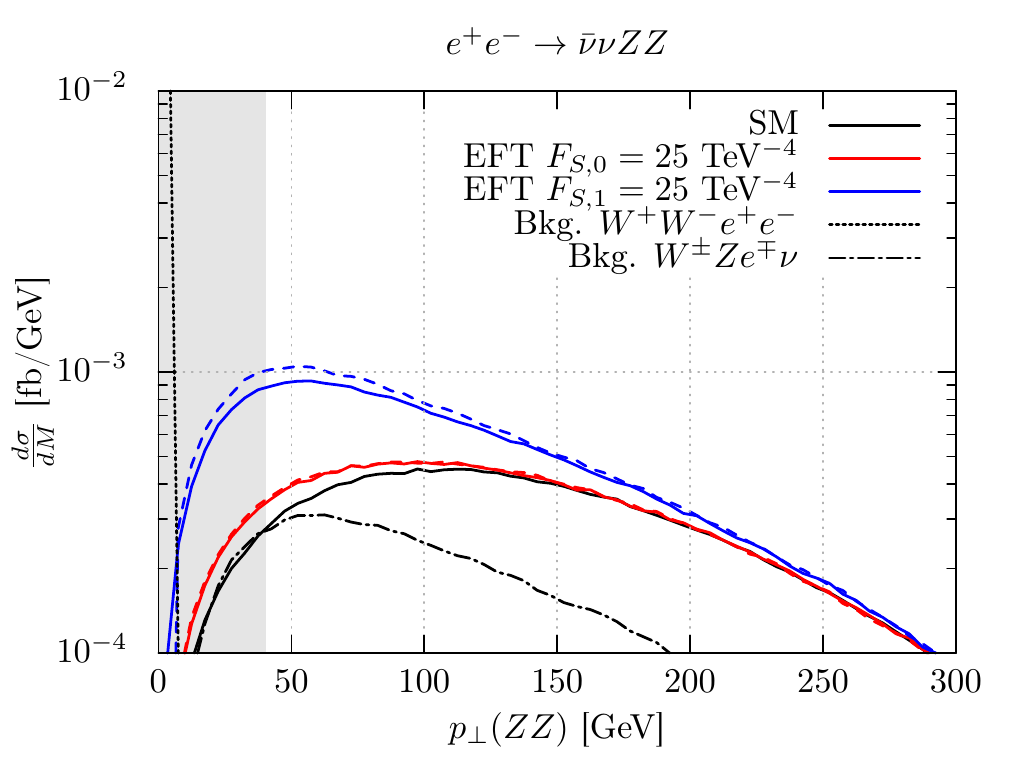}
   \end{subfigure}\\[1pt]%
   \begin{subfigure}[t]{0.46\textwidth}
      \includegraphics[width=\textwidth]{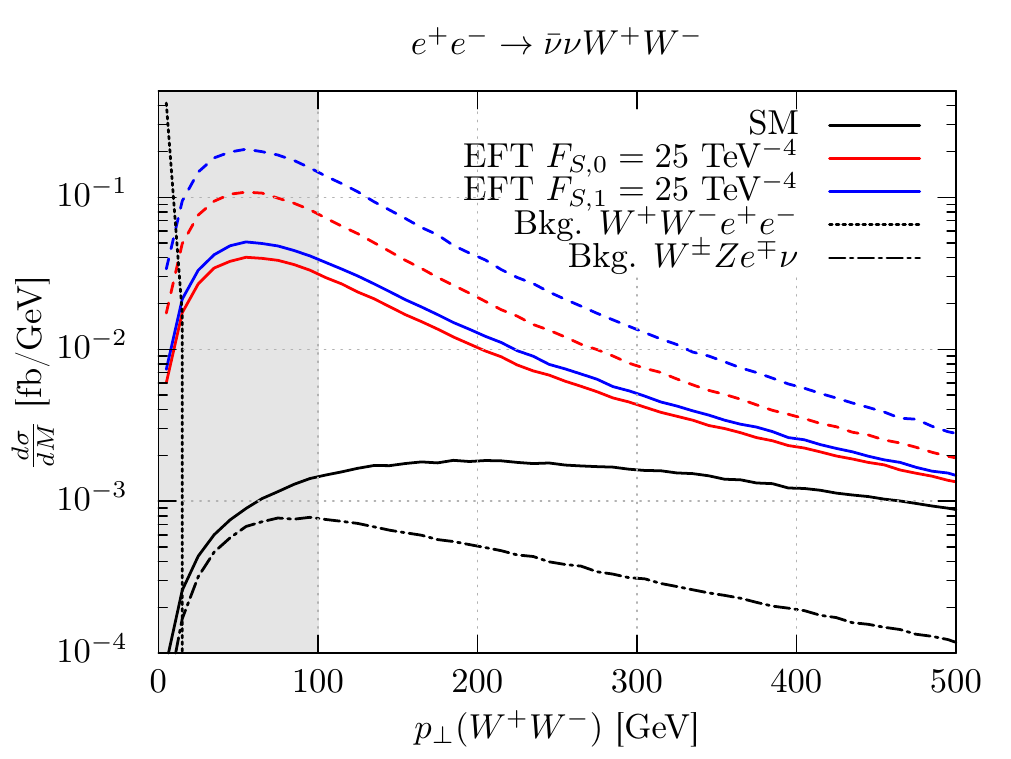}
   \end{subfigure}\hfill%
   \begin{subfigure}[t]{0.46\textwidth}
      \includegraphics[width=\textwidth]{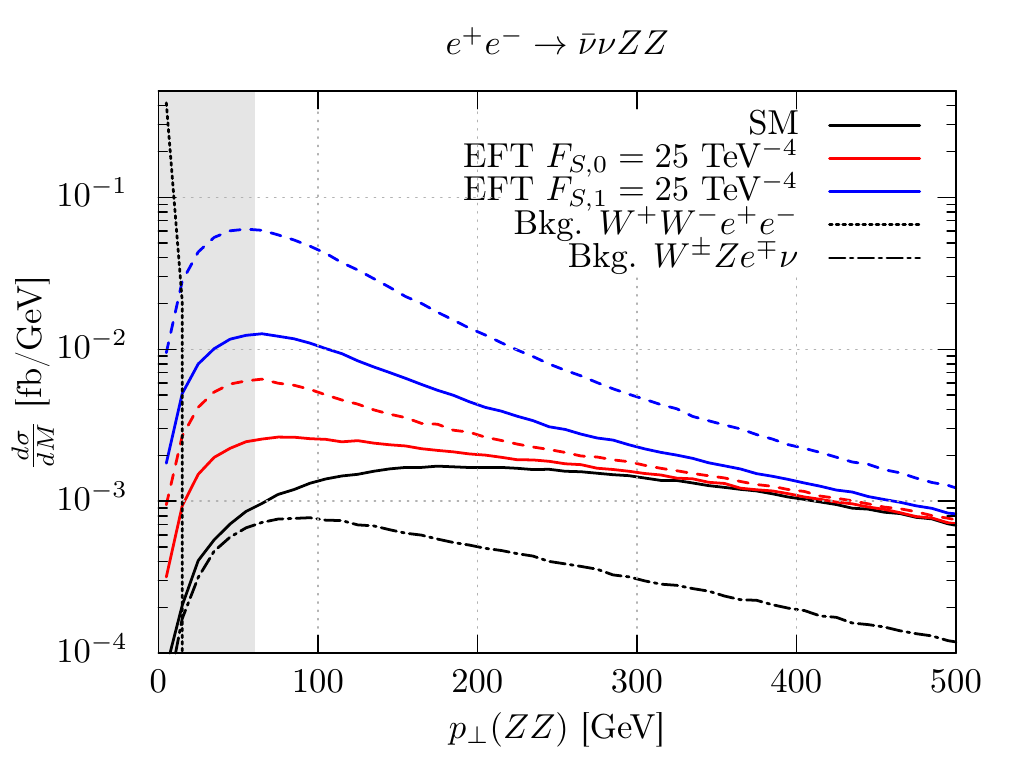}
   \end{subfigure}%
   \caption{Differential cross sections depending on the transverse
            momentum of the $W$ (left plots) and the $Z$
            boson pair (right plots) at center-of-mass energies of
            $1\;\TeV$ (upper plots), $1.4\;\TeV$ (middle plots)
            and $3\;\TeV$ (lower plots).
            The solid lines show the signal process
            $\bar{\nu}\nu W^+W^-(ZZ)$ with SM values
            $F_{S,0}=F_{S,1}=0$. The red/blue lines indicate the
            signal process with non SM value $F_{S,0,1}=25/50\;\TeV\,^{-4}$
            (dashed lines: naive EFT results, solid lines: unitarized
            results). In addition, the two SM background processes
            $W^+W^-e^+ e^-$  and $W^\pm Z e^\mp \nu$ (with 13.6\%
            misidentification probability) are also plotted. The
            shaded area is removed by the cut on the $W(Z)$ boson
            system.  All other cuts have been applied as described.
            No detection  efficiencies are included.}
   \label{i:Pt}
\end{figure}

\begin{figure}[p]
   \begin{subfigure}[t]{0.46\textwidth}
      \includegraphics[width=\textwidth]{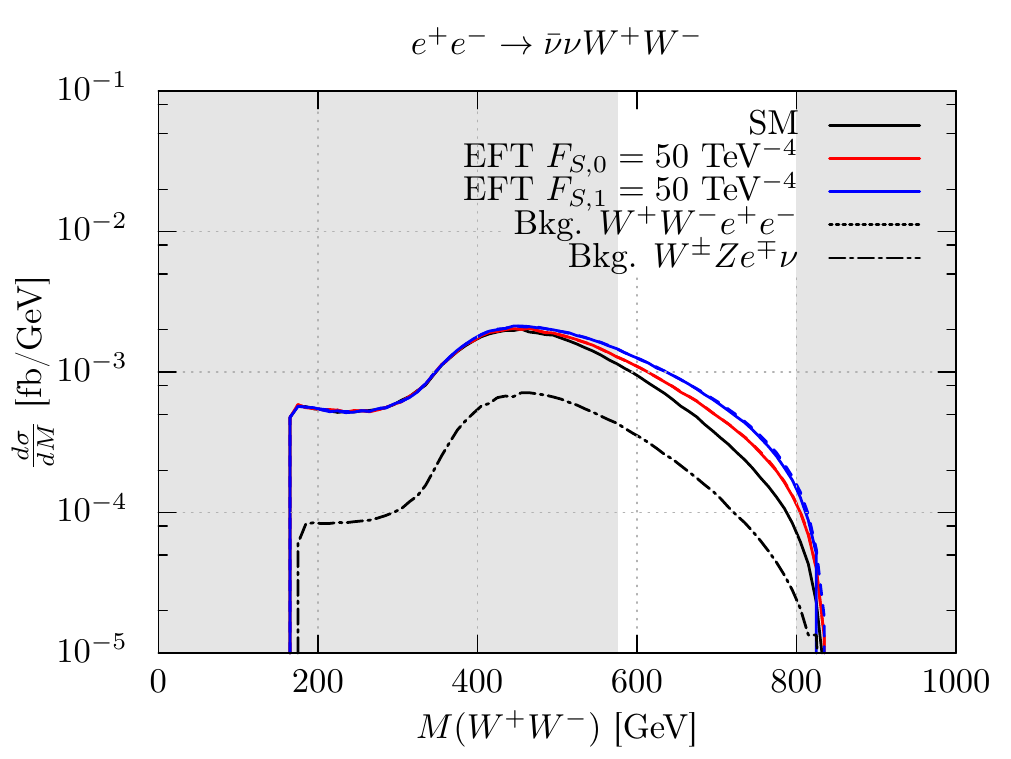}
   \end{subfigure}\hfill%
   \begin{subfigure}[t]{0.46\textwidth}
      \includegraphics[width=\textwidth]{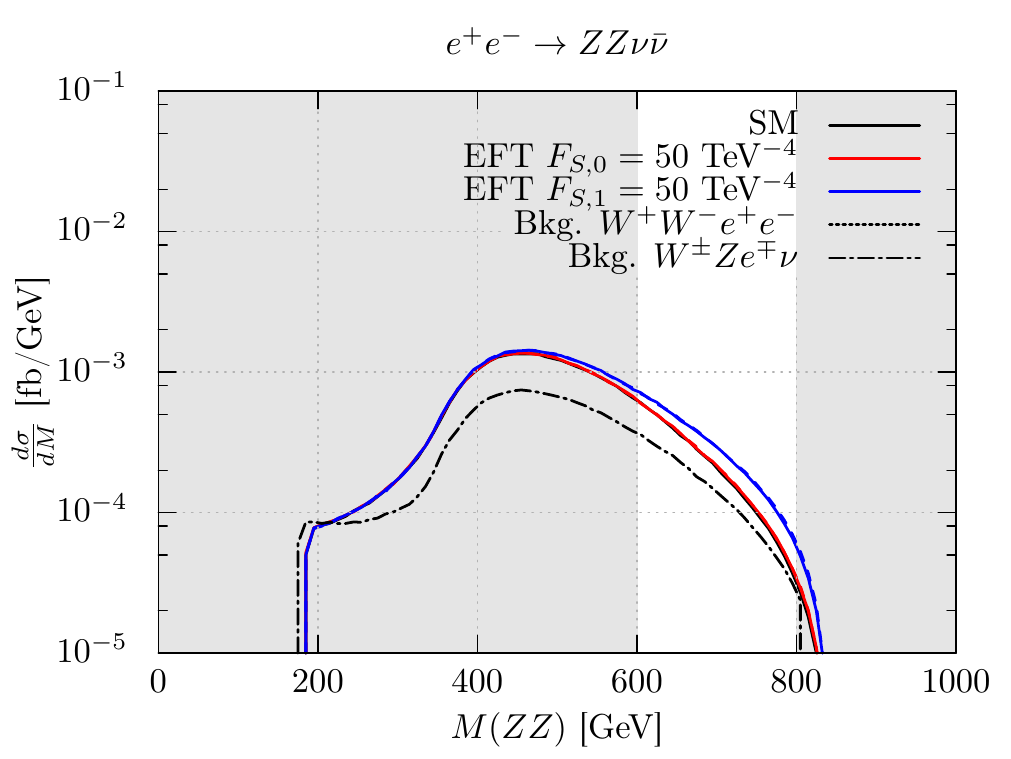}
   \end{subfigure}\\[1pt]%
   \begin{subfigure}[t]{0.46\textwidth}
      \includegraphics[width=\textwidth]{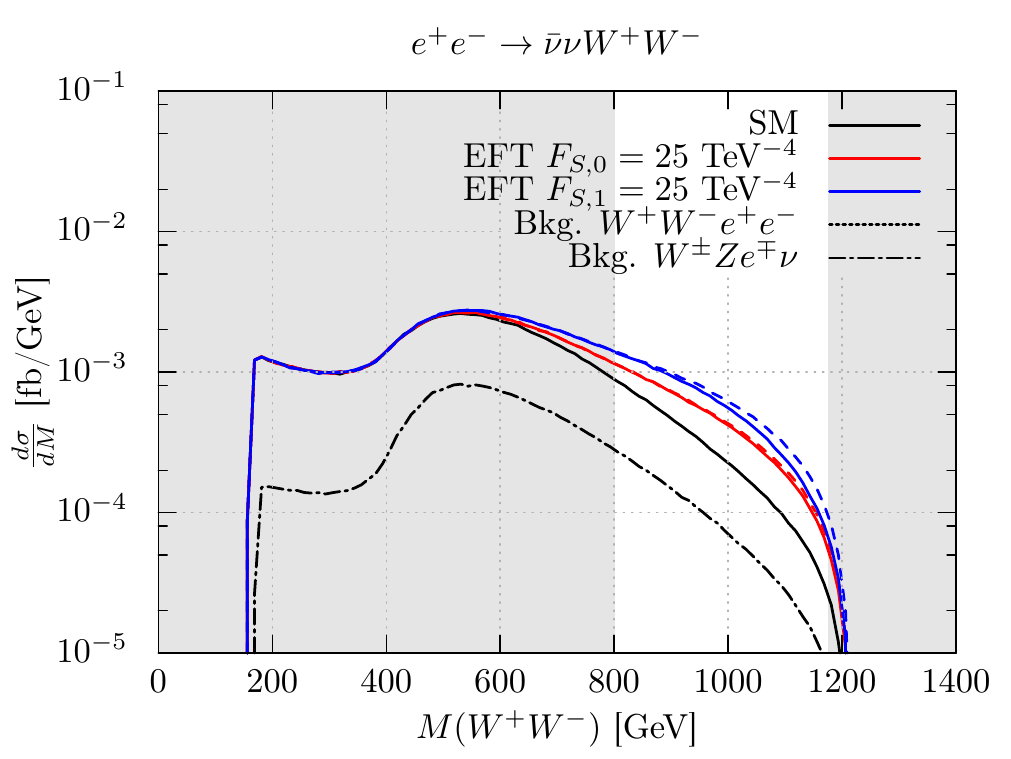}
   \end{subfigure}\hfill%
   \begin{subfigure}[t]{0.46\textwidth}
      \includegraphics[width=\textwidth]{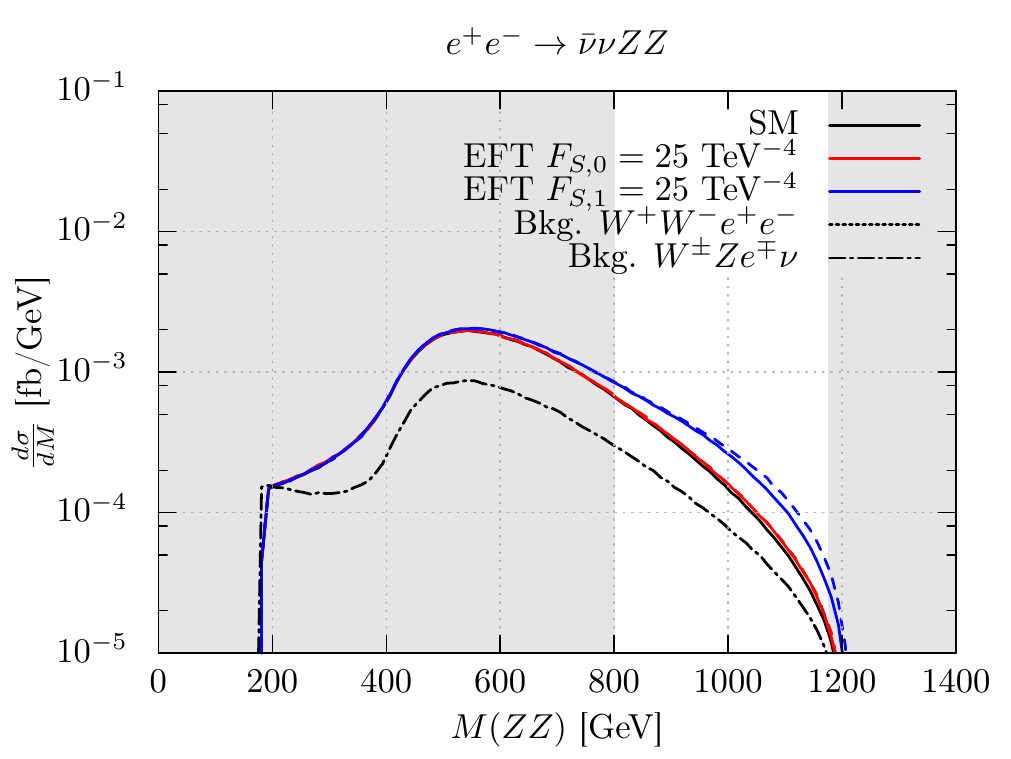}
   \end{subfigure}\\[1pt]%
   \begin{subfigure}[t]{0.46\textwidth}
      \includegraphics[width=\textwidth]{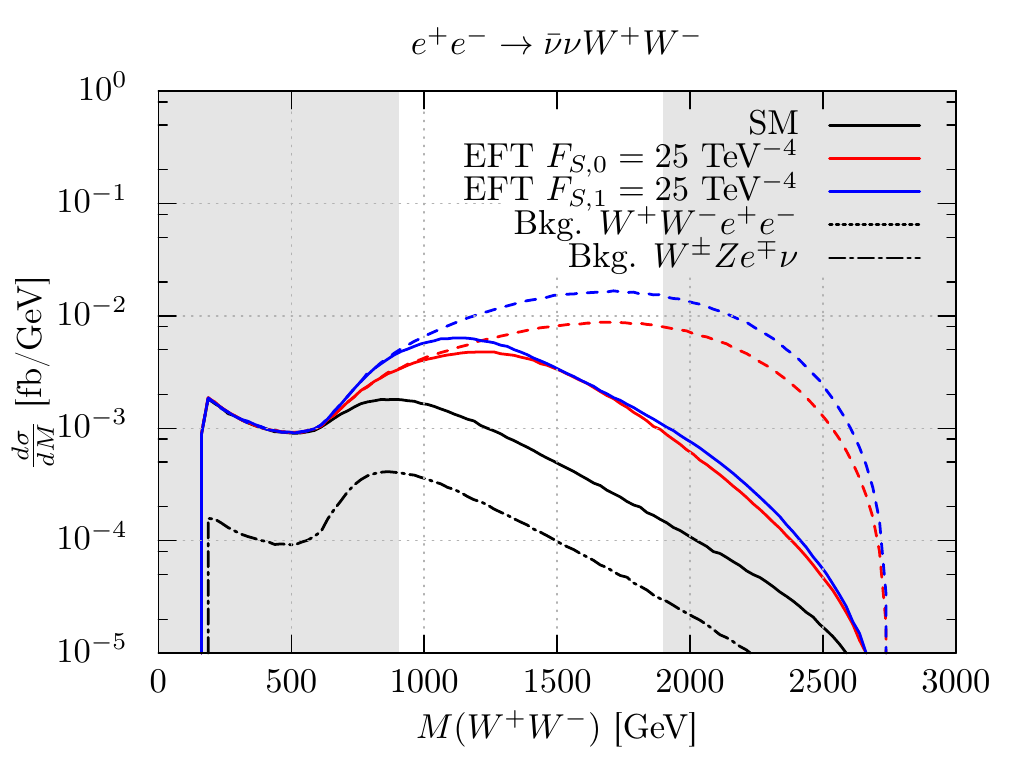}
   \end{subfigure}\hfill%
   \begin{subfigure}[t]{0.46\textwidth}
      \includegraphics[width=\textwidth]{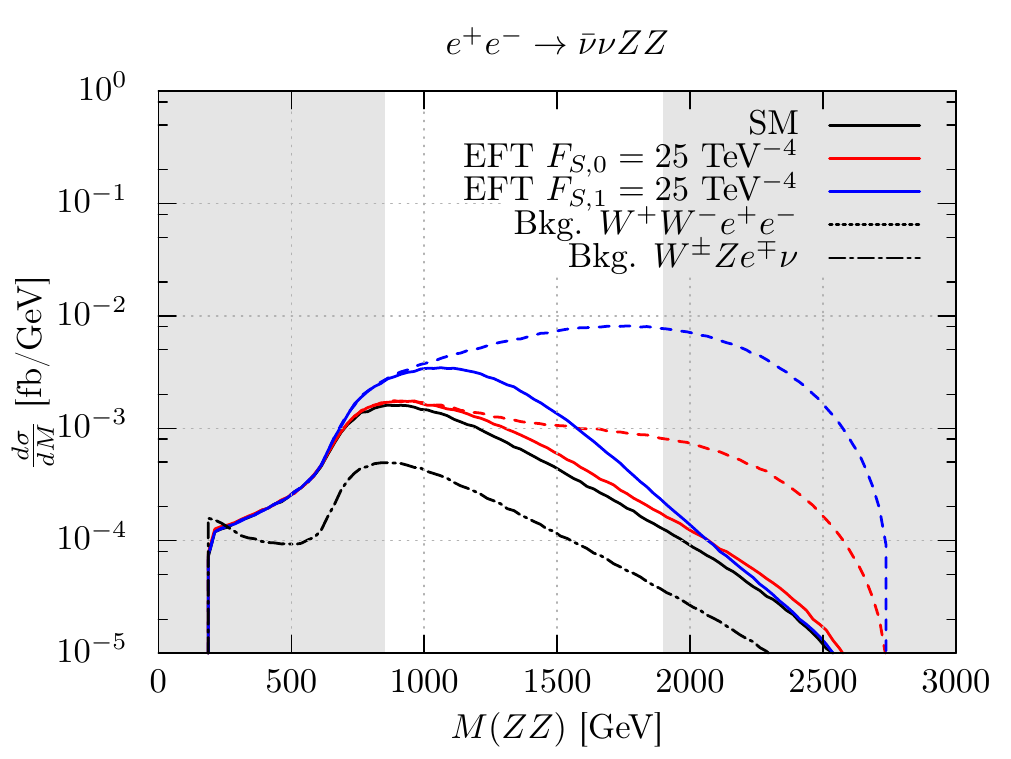}
   \end{subfigure}%
   \caption{Differential cross sections depending on the invariant
            mass of the $W$ (left plots) and the $Z$ boson
            pair (right plots) at center-of-mass energies of
            $1\;\TeV$ (upper plots), $1.4\;\TeV$ (middle plots)
            and $3\;\TeV$ (lower plots).
            The solid lines show the signal process
            $\bar{\nu}\nu W^+W^-(ZZ)$ with SM values
            $F_{S,0}=F_{S,1}=0$. The red/blue lines indicate the
            signal process with non SM value $F_{S,0,1}=25/50\;\TeV\,^{-4}$
            (dashed lines: naive EFT results, solid lines: unitarized
            results). In addition, the two SM background processes
            $W^+W^-e^+ e^-$ and $W^\pm Z e^\mp \nu$ (with 13.6\%
            misidentification probability) are also plotted. The
            shaded area is removed by the cut on the $W(Z)$ boson
            system.  All other cuts have been applied as described.
            No detection  efficiencies are included.}
   \label{i:Minv}
\end{figure}

\begin{figure}[p]
   \begin{subfigure}[t]{0.5\textwidth}
      \includegraphics[width=\textwidth]{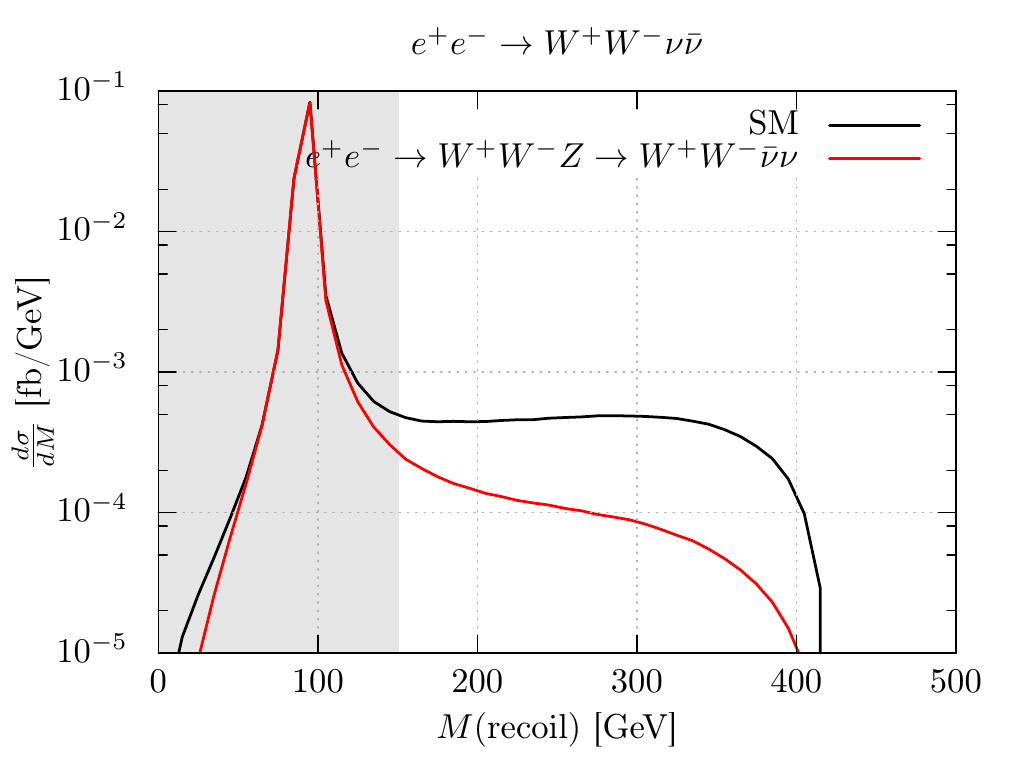}
   \end{subfigure}\hfill%
   \begin{subfigure}[t]{0.5\textwidth}
      \includegraphics[width=\textwidth]{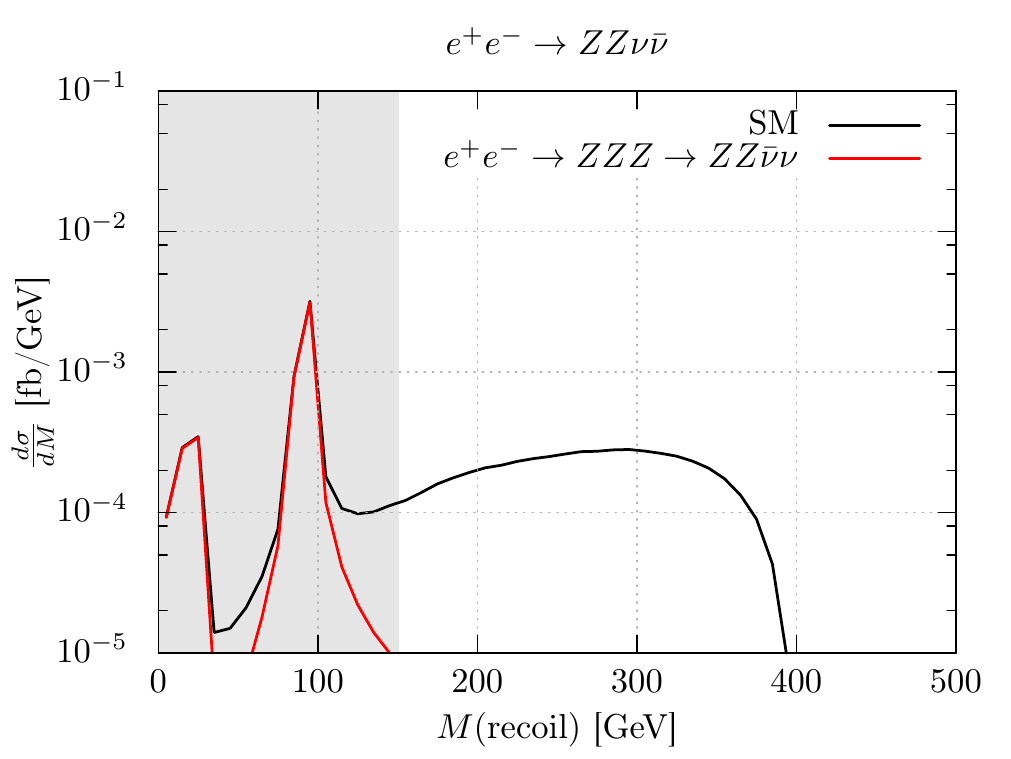}
   \end{subfigure}\\[1pt]%
   \begin{subfigure}[t]{0.5\textwidth}
      \includegraphics[width=\textwidth]{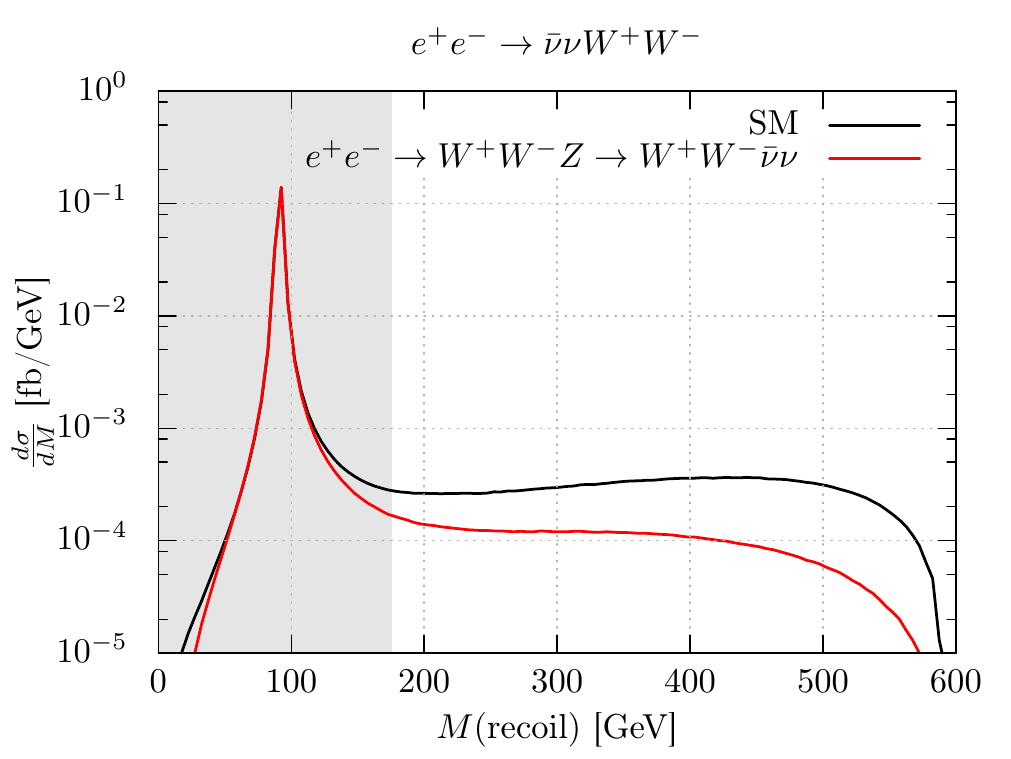}
   \end{subfigure}\hfill%
   \begin{subfigure}[t]{0.5\textwidth}
      \includegraphics[width=\textwidth]{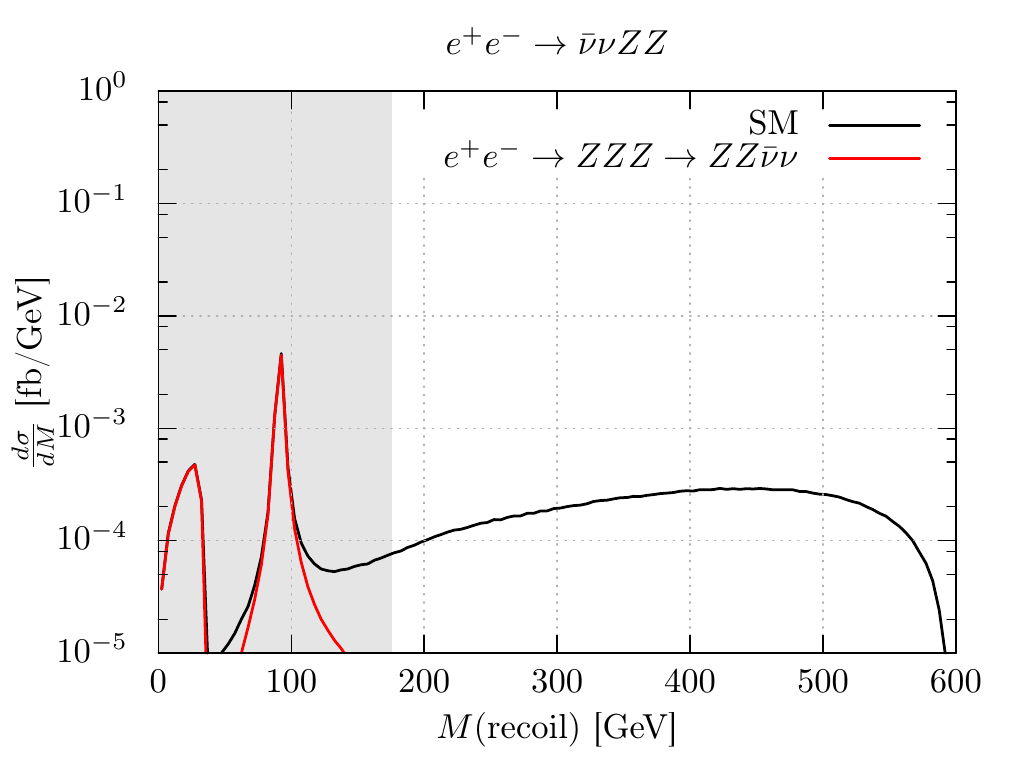}
   \end{subfigure}\\[1pt]%
   \begin{subfigure}[t]{0.5\textwidth}
      \includegraphics[width=\textwidth]{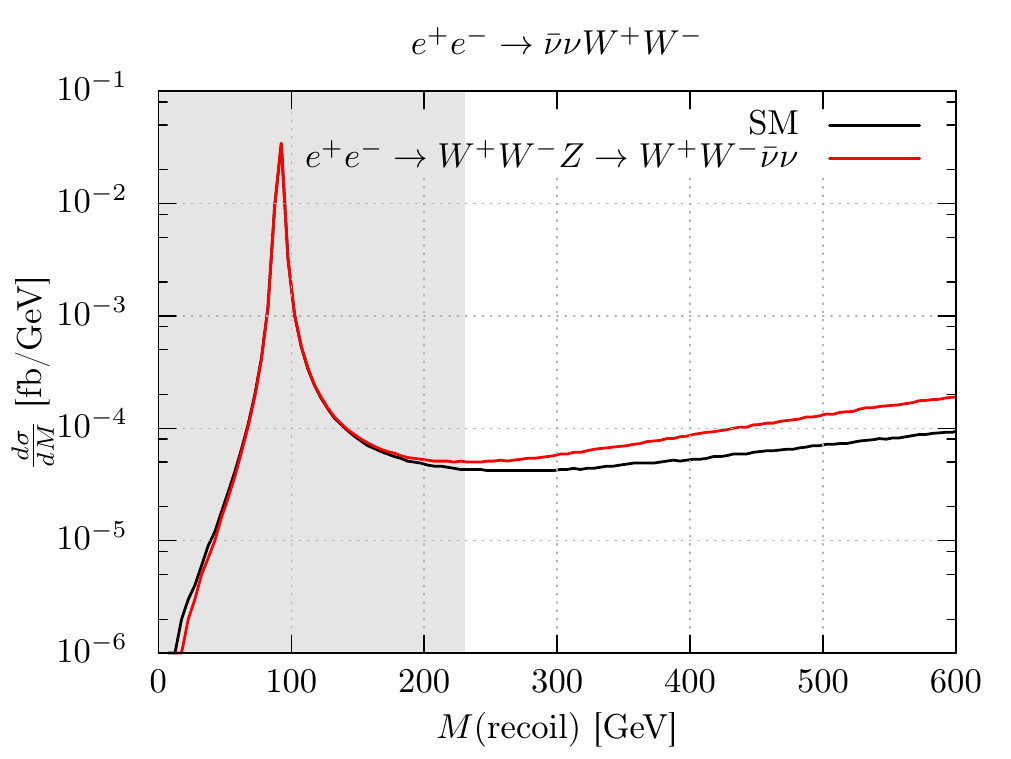}
   \end{subfigure}\hfill%
   \begin{subfigure}[t]{0.5\textwidth}
      \includegraphics[width=\textwidth]{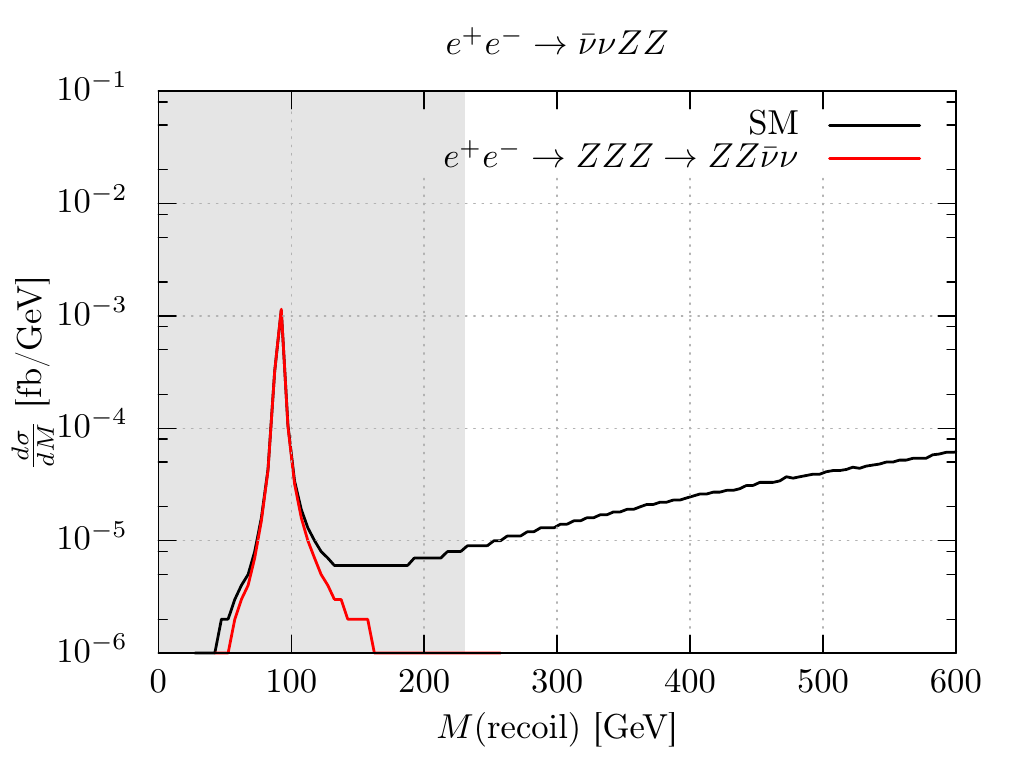}
   \end{subfigure}%
   \caption{Differential cross sections depending on the recoil
            mass of the $W$ (left plots) and the $Z$ boson
            pair (right plots) at center-of-mass energies of
            $1\;\TeV$ (upper plots), $1.4\;\TeV$ (middle plots)
            and $3\;\TeV$ (lower plots).
            The solid lines show the signal process
            $\bar{\nu}\nu W^+W^-(ZZ)$. The red lines show
            the background process $e^+ e^- \to W^+W^-(ZZ)
            Z \to W^+W^-(ZZ)\nu\bar{\nu}$, where the two neutrinos are
            generated through a decaying $Z$ boson. The shaded
            area is removed by the cut on the neutrino system.
            All other cuts have been applied as described above.
            No detection efficiencies are included.}
   \label{i:Rec}
\end{figure}

This expectation is confirmed by the numerical results which we
display in Figs.~\ref{i:Pt}, \ref{i:Minv}, \ref{i:Rec}.  In the
plotted distributions, we choose exemplary values for the model
parameters.  As in the SM results listed above, all distributions are
shown for unpolarized, structureless $e^+e^-$ beams.

\begin{figure}[p]
   \begin{subfigure}[t]{0.5\textwidth}
      \includegraphics[width=\textwidth]{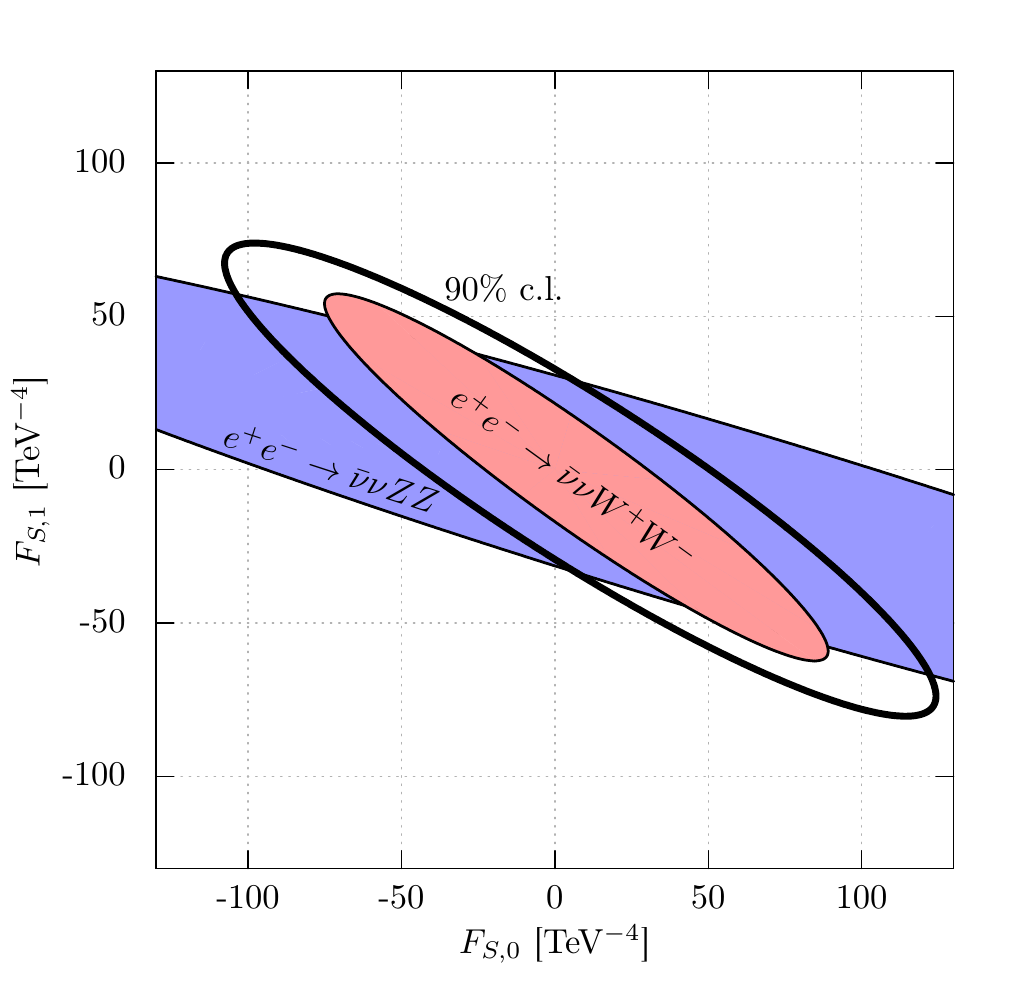}
   \end{subfigure}\hfill%
   \centering
   \begin{subfigure}[t]{0.5\textwidth}
      \includegraphics[width=\textwidth]{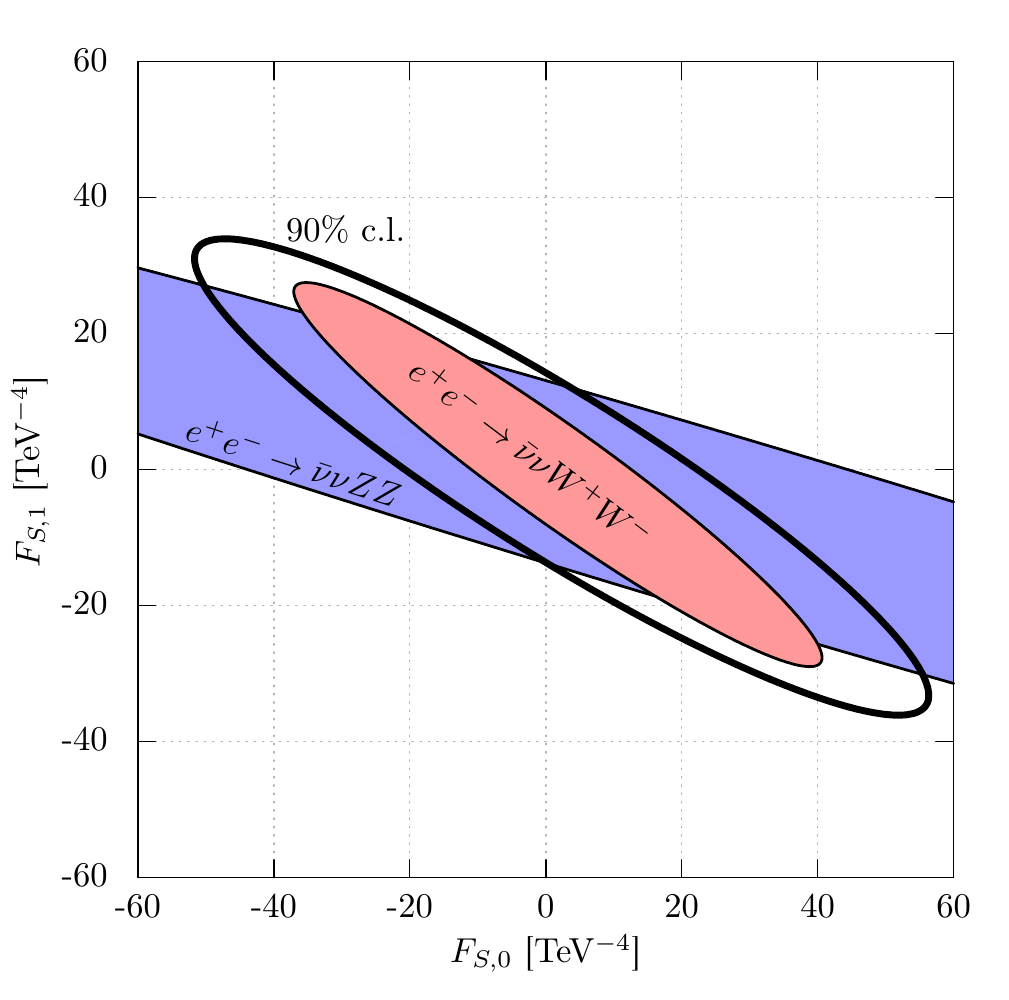}
   \end{subfigure}\hfill%
   \begin{subfigure}[t]{0.5\textwidth}
      \includegraphics[width=\textwidth]{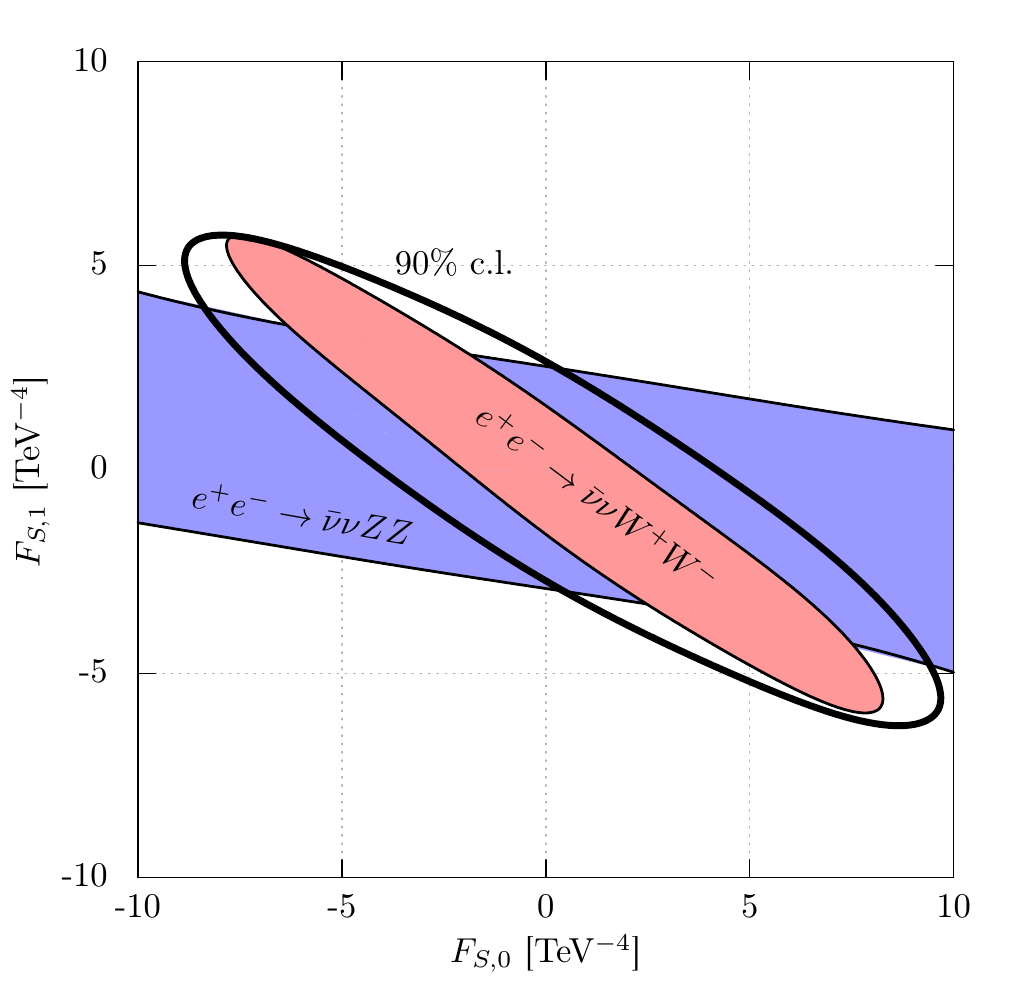}
   \end{subfigure}
   \caption{$\pm 1 \sigma$ exclusion contours in the $F_{S,0} / F_{S,1}$
          plane for the two signal   processes
          $e^+ e^- \to \bar{\nu} \nu W^+ W^-$ and
          $e^+ e^- \to \bar{\nu} \nu ZZ$ including all background
          processes based on the assumption $F_{S,0}=F_{S,1}=0$.
          The $e^- (e^+)$ beam is unpolarized
          at energies of $1.0\;\TeV$ (upper left plot), $1.4\;\TeV$
          (upper right plot) and $3\;\TeV$ (lower plot). The
          corresponding integrated luminosities are $5\;\ab^{-1},
          \ 1.5\;\ab^{-1}$ and $2\;\ab^{-1}$, respectively.
          All cuts have been applied and detection efficiencies
          are included. The thick line indicates the 90\% exclusion
          sensitivities obtained by the combination of the two signal
          channels.
          All cross sections are unitarized.}
   \label{i:plane2}
\end{figure}

The result of this calculation can be expressed in terms of exclusion
contours in a two-dimensional
parameter space, centered on the SM as reference point.  The exclusion
contours are based on the hypothesis
that no deviations from the SM are observed in an experiment.  Given that our
estimates follow from a simple cut-based analysis, we find it
sufficient to calculate a small number of parameter points and
suitably interpolate.
In Fig.~\ref{i:plane2}, we show the results for the
sensitivity to $F_{S,0}$ and $F_{S,1}$.  In addition to the exclusion
contours that we obtain for the $W^+W^-$ and $ZZ$ final states,
respectively, we
indicate a 90\,\% exclusion limit that would be deduced from
combining both channels.  Comparing the three selected collider
energies, we conclude that increasing the energy from $1\;\TeV$ to
$3\;\TeV$ improves the sensitivity by roughly one order of magnitude,
ultimately $\Delta F_{S,0/1}\sim 5\,\TeV^{-4}$.
These results may be compared to the run-I LHC limit on the same parameters,
as obtained by ATLAS~\cite{ATLAS:2014rwa}, $\Delta F_{S,0/1}\sim
500\,\TeV^{-4}$ (cf.~\cite{Sekulla:2015} for the conversion of
exclusion limits).

\begin{figure}[p]
   \begin{subfigure}[t]{0.5\textwidth}
      \includegraphics[width=\textwidth]{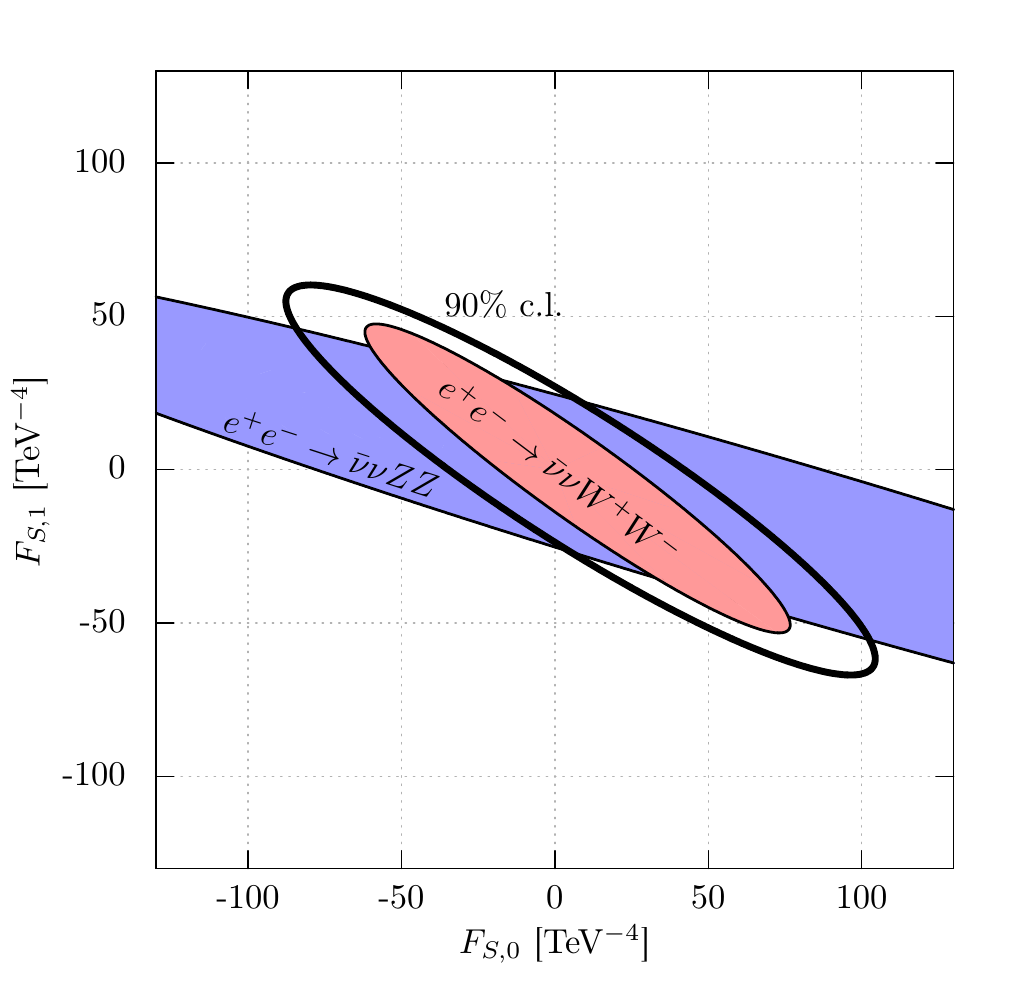}
   \end{subfigure}\hfill%
   \begin{subfigure}[t]{0.5\textwidth}
      \includegraphics[width=\textwidth]{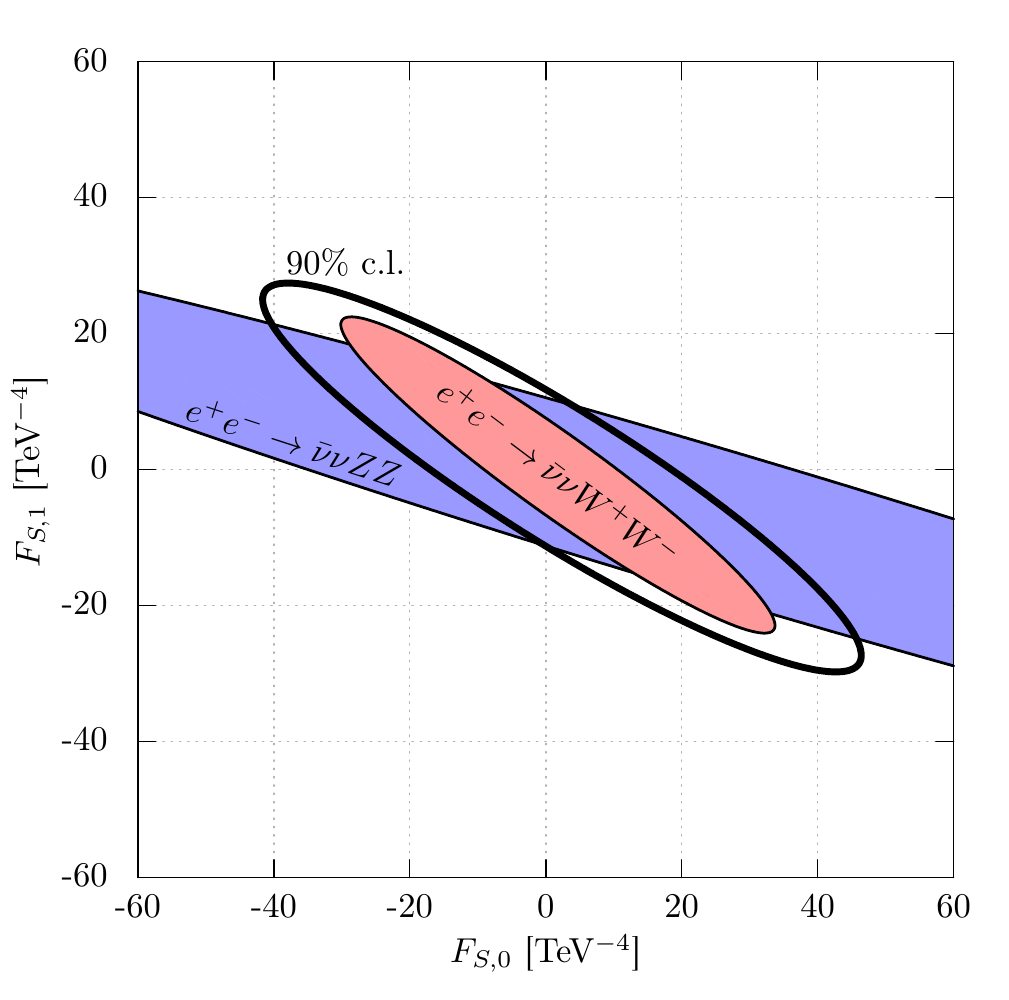}
   \end{subfigure}\hfill%
   \centering
   \begin{subfigure}[t]{0.5\textwidth}
      \includegraphics[width=\textwidth]{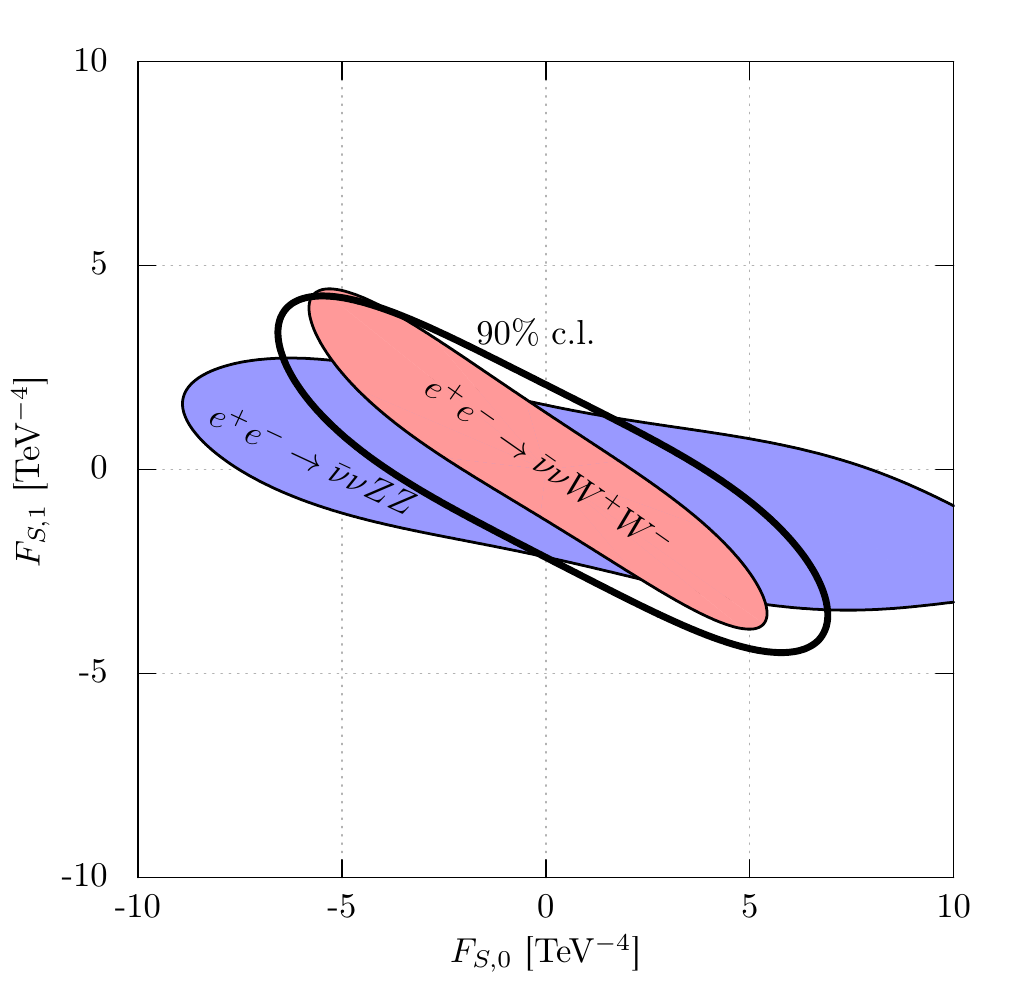}
   \end{subfigure}
   \caption{$\pm 1 \sigma$ exclusion contours in the $F_{S,0} / F_{S,1}$
          plane for the two signal   processes
          $e^+ e^- \to \bar{\nu} \nu W^+ W^-$ and
          $e^+ e^- \to \bar{\nu} \nu ZZ$ including all background
          processes based on the assumption $F_{S,0}=F_{S,1}=0$.
          The $e^- (e^+)$ beam is polarized at a degree of 80\%(0\%)
          at energies of $1.0\;\TeV$ (upper left plot), $1.4\;\TeV$
          (upper right plot) and $3\;\TeV$ (lower plot). The
          corresponding integrated luminosities are $5\;\ab^{-1},
          \ 1.5\;\ab^{-1}$ and $2\;\ab^{-1}$, respectively.
          All cuts have been applied and detection efficiencies
          are included. The thick line indicates the 90\% exclusion
          sensitivities obtained by the combination of the two signal
          channels.
          All cross sections are unitarized.}
   \label{i:plane}
\end{figure}

\begin{figure}[hbt]
 \centering
  \includegraphics[width=0.5\textwidth]{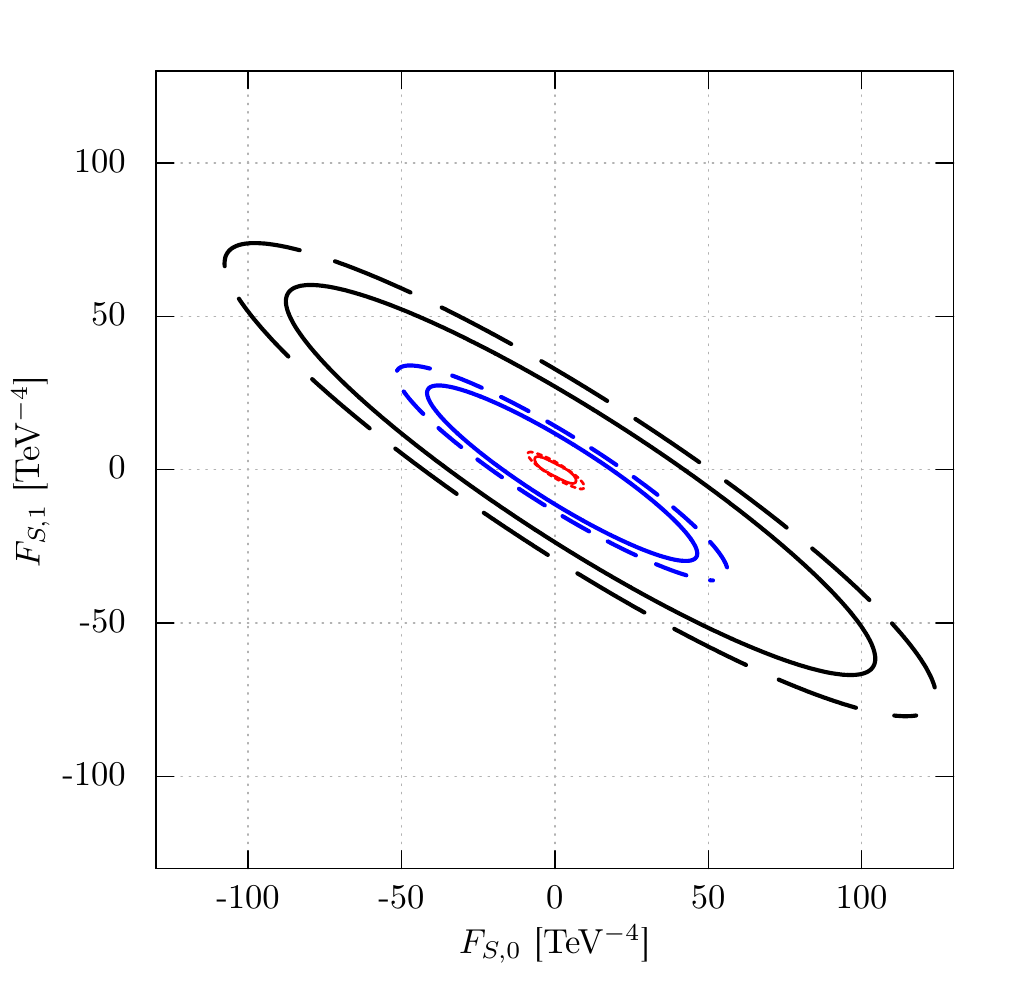}
  \caption{90\;\% exclusion sensitivities for polarized (solid) and
           unpolarized (dashed) particle beams at energies of
           $\sqrt{s}=1 \text{ (black)},\ 1.4 \text{ (blue)},\ 3\,\TeV
           \text{ (red)}$ combined, assuming integrated luminosities
           of $5\;\ab^{-1}, \ 1.5\;\ab^{-1}$ and $2\;\ab^{-1}$,
           respectively. }
  \label{i:alls}
\end{figure}

In Fig.~\ref{i:plane}, we repeat the same analysis for an assumed beam
polarization of 80\,\% ($e^-$), 0\,\% ($e^+$).  The polarization effect
enhances the signal more than the background and thus improves the sensitivity
by another factor of $1.5$.  Finally, Fig.~\ref{i:alls} displays the
same 90\,\% exclusion contours with polarization for all three energy
values in a common plot, for
convenience of the reader.

We deliberately have included only the two channels $W^+W^-$ and $ZZ$.
Additional information can be gained by evaluating the other possible final
states.  However, these channels suffer from a larger fraction of
$\gamma$-induced background and add independent information only if we relax
the custodial-symmetry assumption, such that $W$ and $Z$ states are no longer
related.  For the current analysis, additional channels are of minor importance.

The sensitivity range for the anomalous coupling is of the order
$1/\Lambda_{\text{eff}}^4$, where the effective scale
$\Lambda_{\text{eff}}$ is given by the $W/Z$ pair energy where the
differential distribution is maximal, cf.~Fig.~\ref{i:Minv}.
This is the region where the anomalous couplings
have the strongest impact --- at low energies, their effect is
naturally suppressed, while at high energies, unitarity becomes
saturated and the sensitivity disappears again.  Clearly, the
measurement is limited by statistics.

For the purpose of this study, we rely only on the total cross section
within a fiducial phase space which is isolated by the cut strategy as
described above.  Further details can be deduced from the vector-boson
pair invariant mass distribution, as displayed in the figures.  The
availability of this important observable, in form of the total
hadronic energy and momentum, is a great advantage of the clean
lepton-collider environment, as compared to the situation at a hadron
collider.

\begin{figure}[hbt]
   \begin{subfigure}[t]{0.5\textwidth}
      \includegraphics[width=\textwidth]{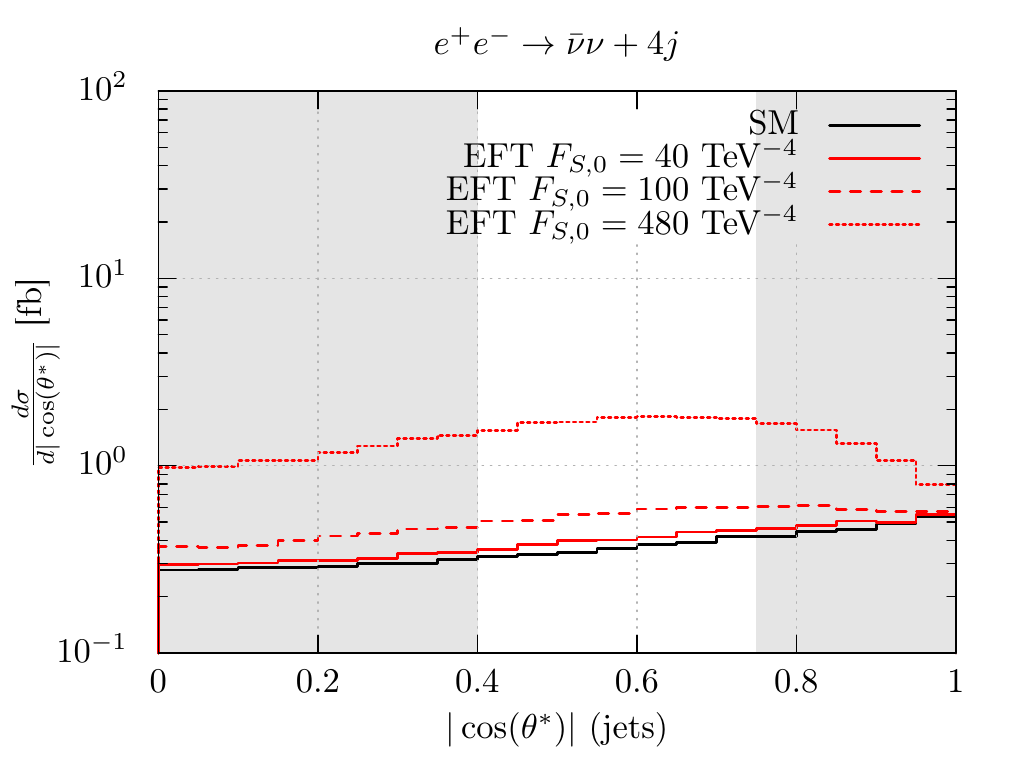}
   \end{subfigure}\hfill%
   \begin{subfigure}[t]{0.5\textwidth}
      \includegraphics[width=\textwidth]{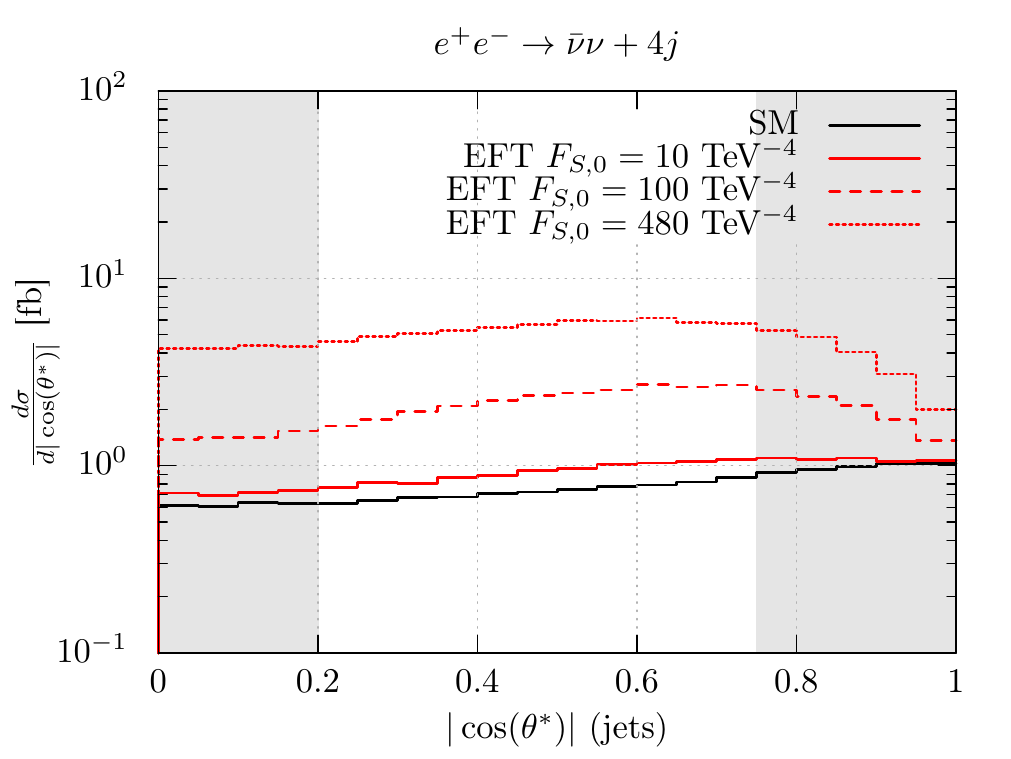}
   \end{subfigure}
   \caption{Differential cross sections of the process
            $e^+e^- \bar{\nu} \to \nu + 4j$ at center-of-mass
            energies of $1.4\;\TeV$ (left plot) and $3\;\TeV$
            (right plot) depending on the jet pairs $|cos(\theta^*)|$
            at different values of $F_{S,0}$. All cuts have been
            applied as described above. No detection efficiencies
            are included and all cross sections are unitarized.}
   \label{i:Theta_star}
\end{figure}

A complete experimental analysis, which is beyond the scope of the
present paper, should take the
simulation results and apply a proper fitting procedure that takes
into account all available information.  Such an analysis of VBS data
would exploit a complete set of observables.  In particular, it will
be advantageous to resolve the decay products of the vector bosons
into individual jets -- quarks in the language of the partonic
elementary process -- and take into account their angular and energy
distributions.  For illustration of the added value, we have generated
WHIZARD event samples for the complete exclusive process $e^+e^-
 \to \bar{\nu} \nu + 4j$ with all possible Feynman graphs included,
summed over neutrino and quark flavors, for the SM and for some
nonzero values of the EFT operator coefficient $F_{S,0}$,
Fig.~\ref{i:Theta_star}.

In this figure, we show the distribution in the polar angle
$\theta^*$ between the final state jets in the rest
frame of the parent (off-shell) vector boson. This cut is applied at
Monte Carlo truth level to both jet pairs including all
combinatorics. Expanding on this result, to
enhance the vector boson scattering signal further above the
background, the following cut on the angle $\theta^*$ could be used
(as before: first number applies for $3\;\TeV$ and the number in
brackets for $1.4\;\TeV$):
\begin{equation}
  0.2(0.4) < |\cos(\theta^*)| < 0.75
\end{equation}
A more sophisticated analysis would exploit the
complete information from this distribution and add in any further
observable that provides discriminating power.

\section{Results for Resonances}

Besides the extrapolated EFT of the previous section, we now consider a
second set of models for
VBS at high energy.  Starting again from the Lagrangian~(\ref{SM-L}),
we add a field with high mass, definite $SU(2)_L\times
SU(2)_R$ quantum numbers and with a minimal coupling to a
Higgs-Goldstone current in the Lagrangian, as a possible resonance.  For
concreteness, we adopt
resonance quantum numbers and parameters that we did study for the LHC
in Ref.~\cite{Kilian:2015opv}, and specifically select three models,
namely an
isoscalar-scalar $\sigma$,
an isotensor-scalar $\phi$, and an isoscalar-tensor $f$.  For each
model, we introduce the resonance mass and the coupling to the Higgs
current as two independent new parameters.  Ignoring further decay
channels, the coupling and mass also determine the width of the
resonance.

The Lagrangians for the three fields $\sigma$, $f$, $\phi$ are
\begin{subequations}
\label{eq:appendix_lagrangian}
\begin{align}
  \LL_\sigma =&
  \frac{1}{2} \partial_\mu \sigma \partial^\mu \sigma
  -\frac{1}{2} m_\sigma^2 \sigma^2
  + \sigma J_\sigma \, \\
  \LL_\phi =& \frac{1}{2}\sum_{i=s,v,t} \tr{\partial_\mu \Phi_i \partial^\mu \Phi_i
    -m_\Phi^2 \Phi_i^2}
  +\tr{\left(\Phi_t + \frac{1}{2}\Phi_v - \frac{2}{5}\Phi_s \right)J_\phi }\, ,\\
  \LL_f =&\frac{1}{2} \partial_\alpha f_{\mu\nu}\partial^\alpha f^{\mu\nu}
      - \frac{1}{2} m^2 f_{\mu \nu}f^{\mu \nu} \notag\\
     & \;- \partial^\alpha f_{\alpha\mu} \partial_\beta f^{\beta\mu}
      - f^\alpha_{\;\alpha}\partial^\mu\partial^\nu f_{\mu\nu}
      - \frac{1}{2} \partial_\alpha f^\mu_{\;\mu}\partial^\alpha f^\nu_{\;\nu}
      + \frac{1}{2} m^2 f^\mu_{\;\mu} f^\nu_{\;\nu}
      + f_{\mu\nu}J_f^{\mu\nu},
\end{align}
\end{subequations}
where $J_\sigma,J_\phi,J_f$ are the currents which couple to the new fields,
respectively.  We note that the isotensor-scalar $\phi$, defined by its
$SU(2)_R \times SU(2)_L$ quantum numbers $\mathbf{1} \times
\mathbf{1}$, decomposes after electroweak symmetry breaking  into
an isotensor $\Phi_t$,
an isovector $\Phi_v$ and isoscalar $\Phi_s$ transforming under
custodial $SU(2)$.

The currents that interact with the resonances are given by
\begin{subequations}
  \begin{alignat}{2}
    J_{\sigma} &= F_\sigma&&
    \tr{ \left ( \vD_\mu \vH \right )^\dagger \vD^\mu \vH} \, ,\\
    J_{\phi} &= F_\phi&&
    \left (
      \left ( \vD_\mu \vH \right )^\dagger \otimes \vD^\mu \vH
      +\frac {1}{8} \tr{\left ( \vD_\mu \vH \right )^\dagger \vD^\mu \vH }
    \right )\tau^{a} \otimes \tau^a \, , \\
    J^{\mu \nu}_f&=
    F_f &&\left (
      \tr{ \left ( \vD^\mu \vH \right )^\dagger \vD^\nu \vH}
      - \frac{c_f}{4} g^{\mu \nu}
      \tr{ \left ( \vD_\rho \vH \right )^\dagger \vD^\rho \vH}
    \right)  \, .
  \end{alignat}
\end{subequations}
For further details, cf.~\cite{Kilian:2015opv}.

At low energy, each of these three distinct models reduces to a
one-parameter effective theory, since a single combination of mass and
coupling parameters enters into effective values of $F_{S,0}$ and
$F_{S,1}$.  These models therefore provide high-energy
extensions
of the generic low-energy EFT, as alternatives to the
straightforward extrapolation of the preceding section.
The resonance models exhibit more structure than the simple
extrapolation.  All models require unitarization,
which we again implement using the T-matrix framework, since the
interaction operators introduce terms that rise
with energy.

If the effect of the new states is to be sizable and
thus observable, the interactions should be rather strong, so
generically we do
not expect a renormalizable model with tree-level unitary asymptotics.
Nevertheless, weakly interacting models such as a UV-complete
Higgs-singlet model
are included in this model space.  For a renormalizable model, any
terms that apparently rise with energy would cancel against
higher-order contributions.  From a phenomenological perspective,
renormalizable
models are exceptional points in a larger parameter space that
we cover by the above definitions.

\begin{figure}[p]
   \begin{subfigure}[t]{0.5\textwidth}
      \includegraphics[width=\textwidth]{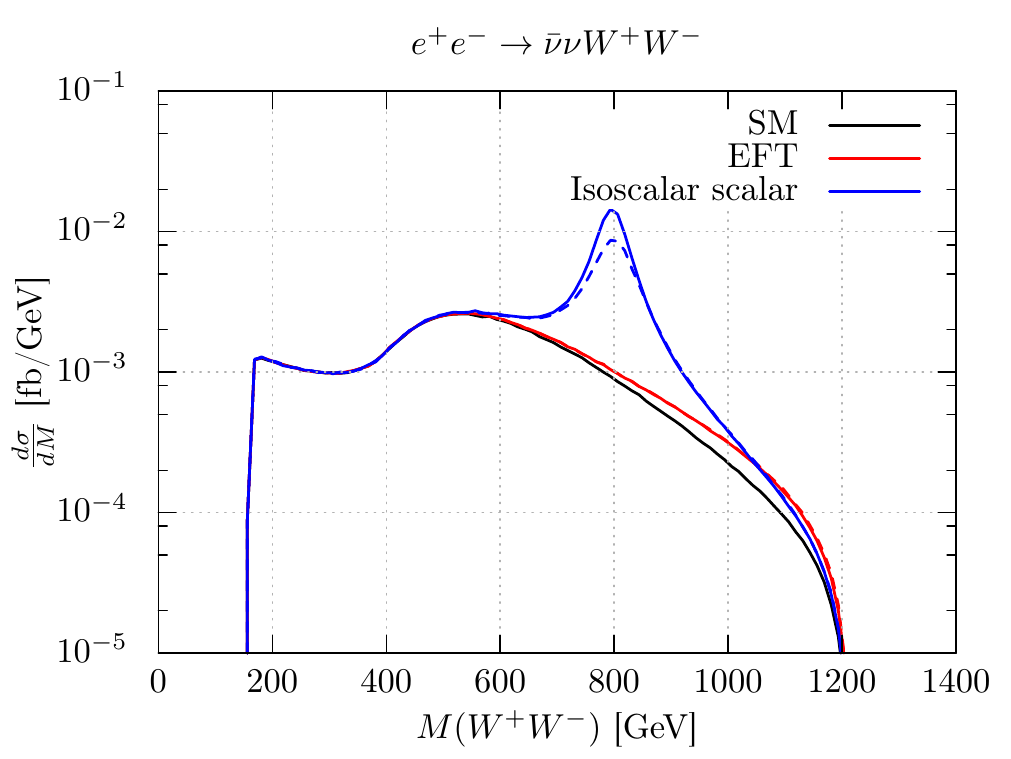}
   \end{subfigure}\hfill%
   \begin{subfigure}[t]{0.5\textwidth}
      \includegraphics[width=\textwidth]{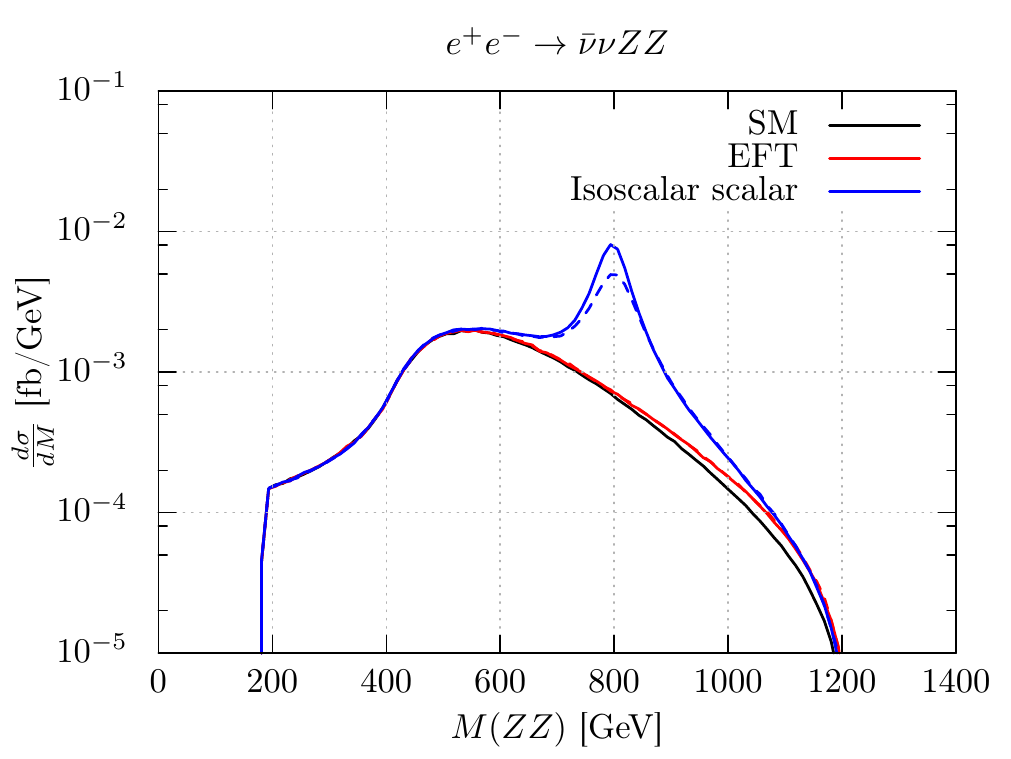}
   \end{subfigure}\\[5pt]%
   \centering
   \begin{subfigure}[t]{0.5\textwidth}
      \includegraphics[width=\textwidth]{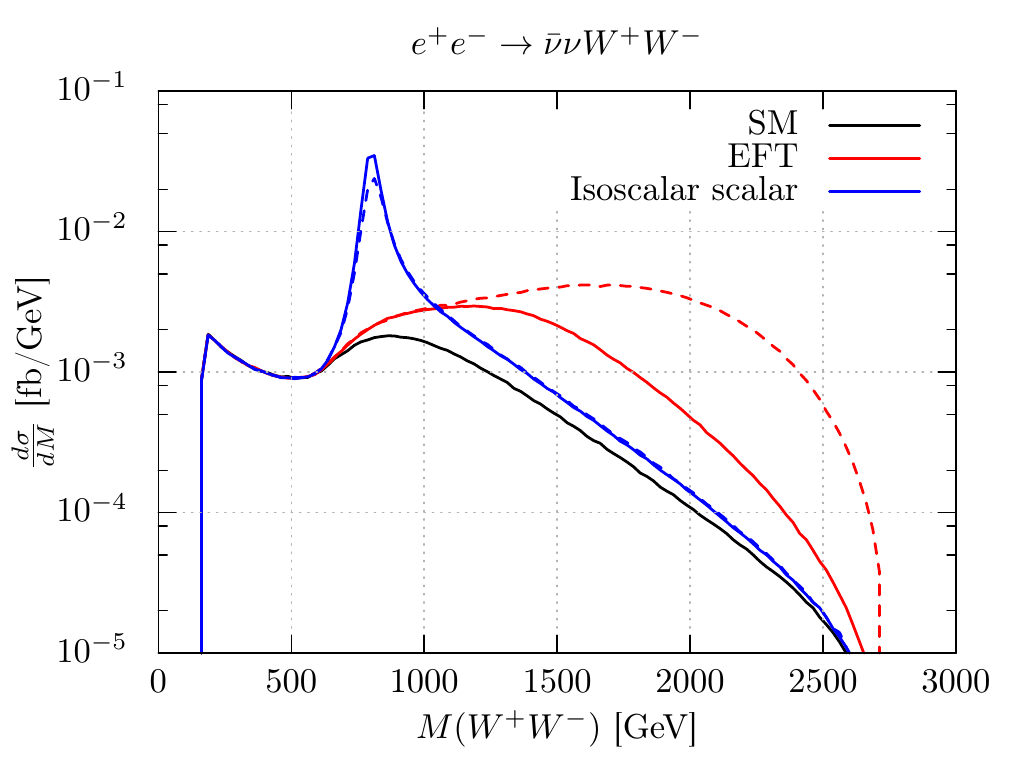}
   \end{subfigure}\hfill%
   \begin{subfigure}[t]{0.5\textwidth}
      \includegraphics[width=\textwidth]{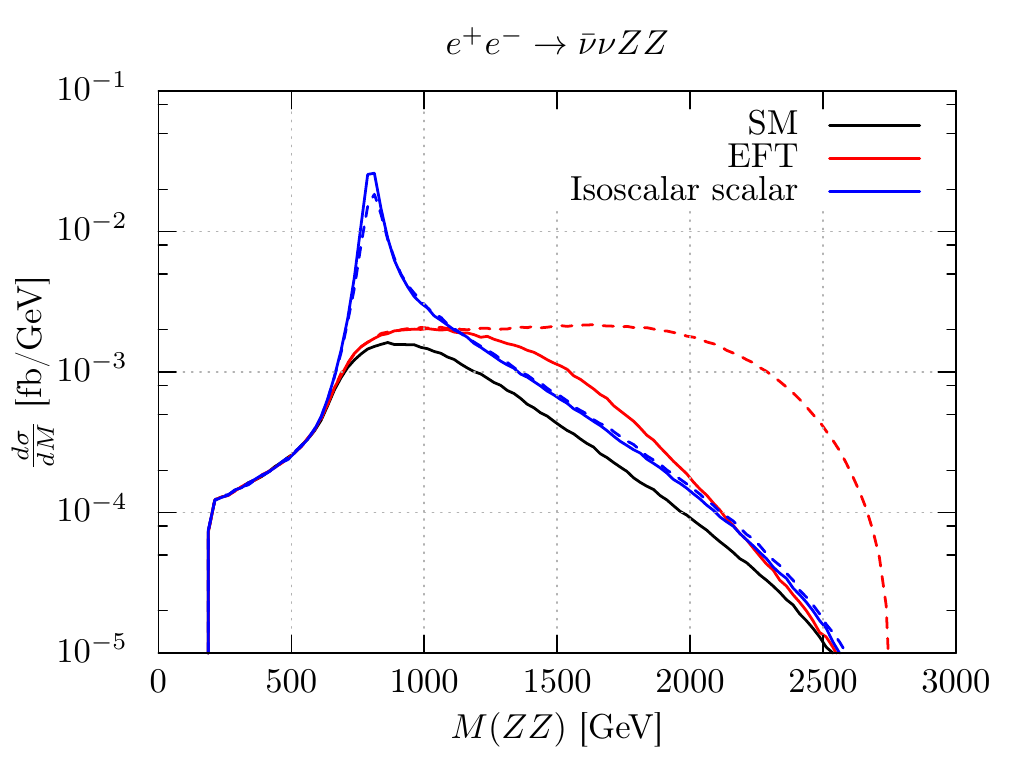}
   \end{subfigure}
   \caption{Differential cross sections including a weakly coupled isoscalar
            scalar resonance (${m_\sigma=800\;\GeV}$,
            $F_\sigma=4.0\;\TeV^{-1}$, $\Gamma_\sigma=80\;\GeV$)
            depending on the invariant mass of the vector boson pair
            at center-of-mass energies of $\sqrt{s}=1.4\;\TeV$
            (upper plots) and $\sqrt{s}=3\;\TeV$ (lower plots).
            Plots on the left show the process
            $e^+e^-\to \bar{\nu}\nu W^+W^-$, plots on the right
            $e^+e^-\to \bar{\nu}\nu ZZ$. Blue line: isoscalar scalar
            resonance, red line: matched EFT results ($F_{S,0}=0$,
            $F_{S,1}=12.3\;\TeV^{-4}$). Solid line:
            unitarized results, dashed line: naive results.}
   \label{i:isoscalar_scalar_1}
\end{figure}

\begin{figure}[p]
   \begin{subfigure}[t]{0.5\textwidth}
      \includegraphics[width=\textwidth]{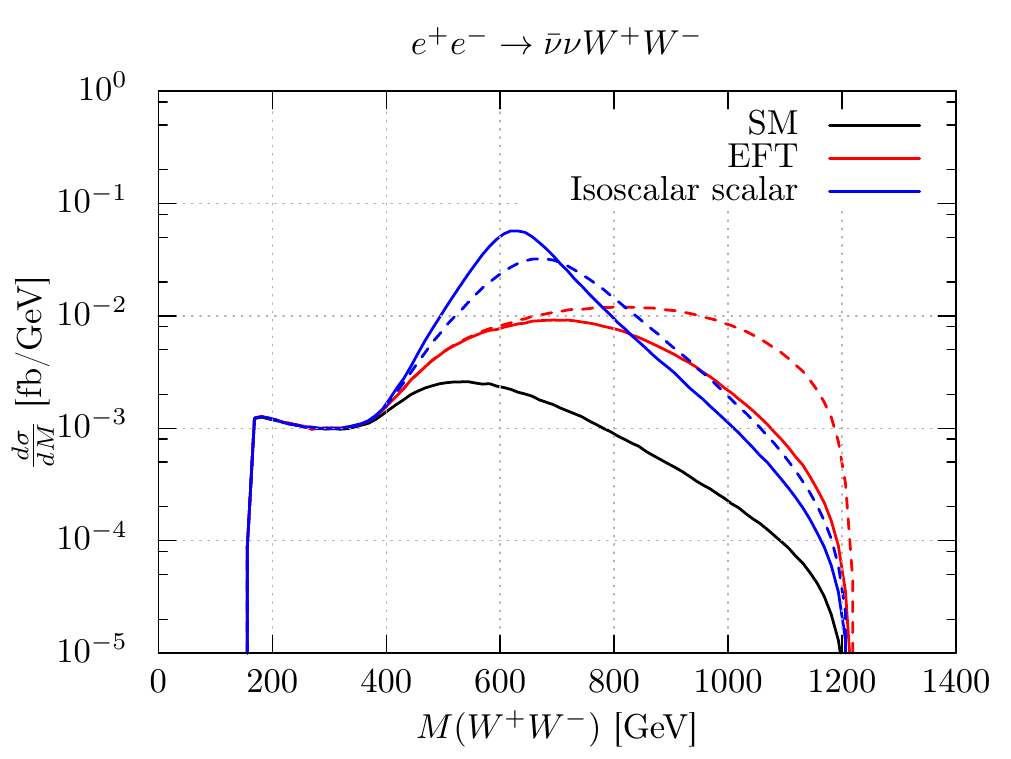}
   \end{subfigure}\hfill%
   \begin{subfigure}[t]{0.5\textwidth}
      \includegraphics[width=\textwidth]{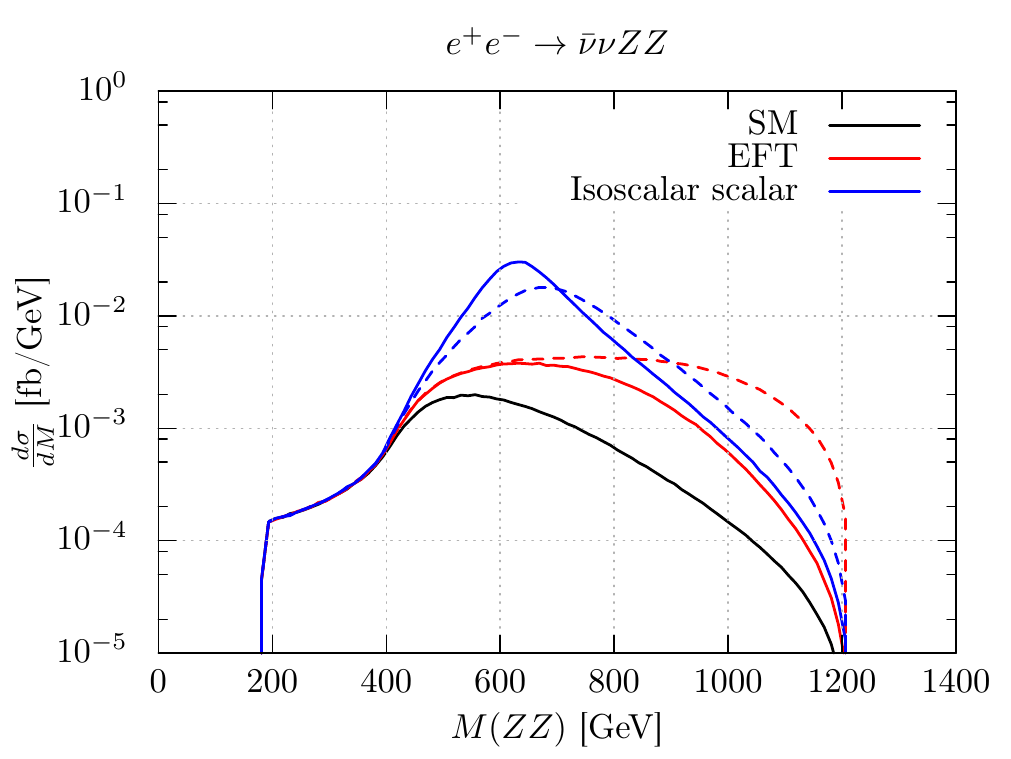}
   \end{subfigure}\\[5pt]%
   \centering
   \begin{subfigure}[t]{0.5\textwidth}
      \includegraphics[width=\textwidth]{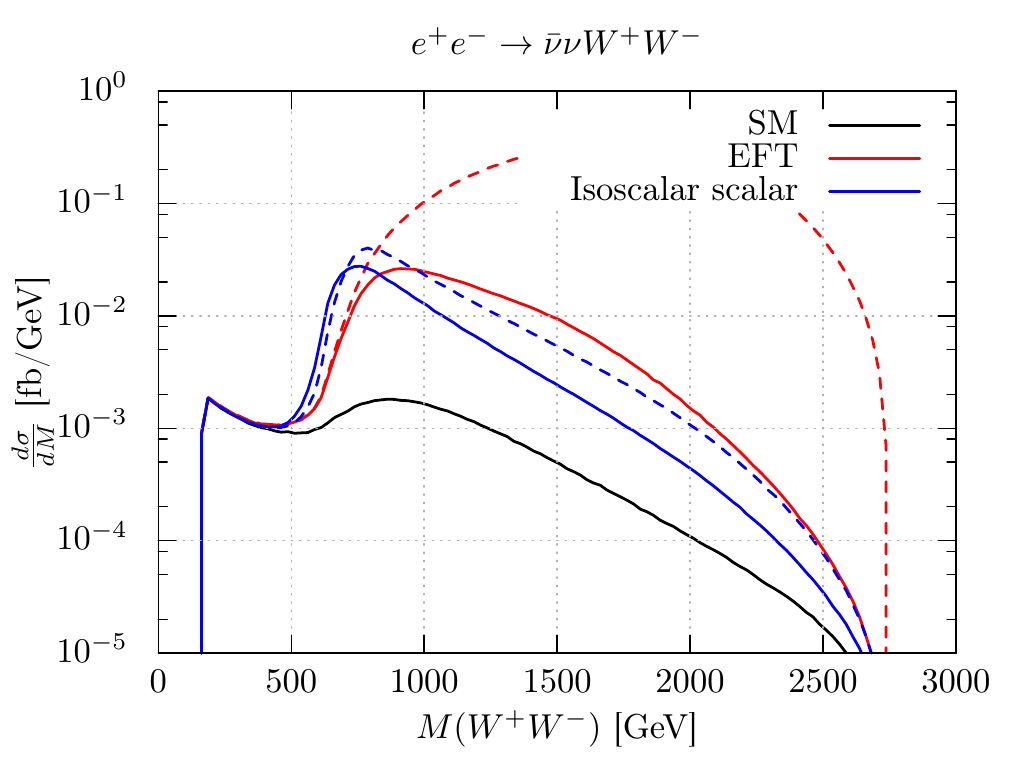}
   \end{subfigure}\hfill%
   \begin{subfigure}[t]{0.5\textwidth}
      \includegraphics[width=\textwidth]{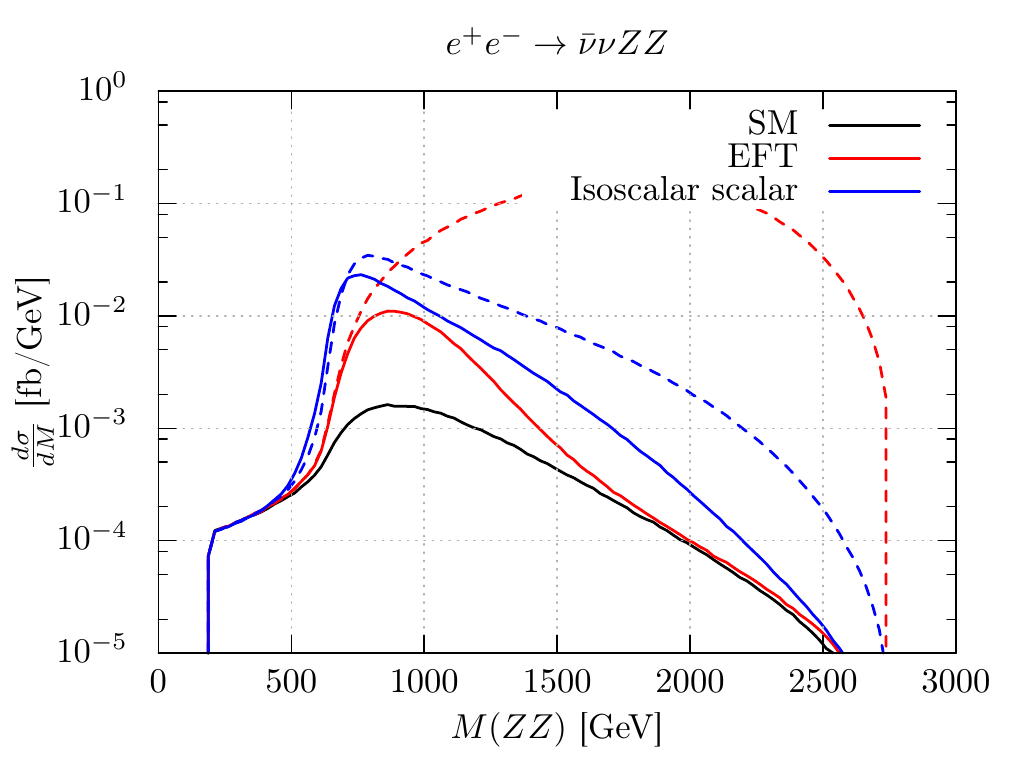}
   \end{subfigure}
   \caption{Differential cross sections including a low lying isoscalar
            scalar resonance (${m_\sigma=650\;\GeV}$,
            $F_\sigma=9.8\;\TeV^{-1}$, $\Gamma_\sigma=260\;\GeV$)
            depending on the invariant mass of the vector boson pair
            at center-of-mass energies of $\sqrt{s}=1.4\;\TeV$
            (upper plots) and $\sqrt{s}=3\;\TeV$ (lower plots).
            Plots on the left show the process
            $e^+e^-\to \bar{\nu}\nu W^+W^-$, plots on the right
            $e^+e^-\to \bar{\nu}\nu ZZ$. Blue line: isoscalar scalar
            resonance, red line: matched EFT results ($F_{S,0}=0$,
            $F_{S,1}=112.6\;\TeV^{-4}$). Solid line:
            unitarized results, dashed line: naive results.}
   \label{i:isoscalar_scalar_2}
\end{figure}

In Fig.~\ref{i:isoscalar_scalar_2}, we display the vector-boson pair
invariant mass distribution for the SM with an additional isoscalar
scalar resonance~$\sigma$, equivalent to an extra Higgs-like singlet boson.
The plots show both the $W^+W^-$ final state (left column) and the
$ZZ$ final state (right column).  We have chosen a moderately high
mass of $M_\sigma=800\;\GeV$ and a rather small width of $\Gamma_\sigma=80\;\GeV$ (blue
curve).  The distribution, which for the full process translates into
the invariant mass of the hadronic (four-jet) system, shows an
unambiguous peak at the resonance mass that is distinguishable in
shape from the SM background (black), given sufficient luminosity.
The peak is more pronounced for $3\;\TeV$ collider energy, but also
clearly visible for $1.4\;\TeV$.  It is evident that for this choice
of parameters, the formally correct EFT expansion (red curve) does not
aproximate the actual model behavior at all.  Unitarization does not
play a significant role except for the naively extrapolated EFT
(dashed red) which overshoots the unitarized curves (solid), and thus the
unitarity bounds, by a substantial amount.

The plots in Fig.~\ref{i:isoscalar_scalar_2} depict a somewhat lighter
isoscalar-scalar resonance ($M_\sigma=650\;\GeV$) with a larger width
($\Gamma_\sigma=260\;\GeV$).  In this case, the resonance shape follows the
shape of the extrapolated EFT but leads to at significantly larger
peak cross section.  Since the equivalent EFT parameter, matching to
the slope of the resonance curve below threshold, is twice as large as
in the previous set of plots, the unphysical behavior is evident that
we would get without unitarization (dashed curves).  This applies not
just to the EFT approximation (red), but also to the resonance curves
(blue).

\begin{figure}[p]
   \begin{subfigure}[t]{0.5\textwidth}
      \includegraphics[width=\textwidth]{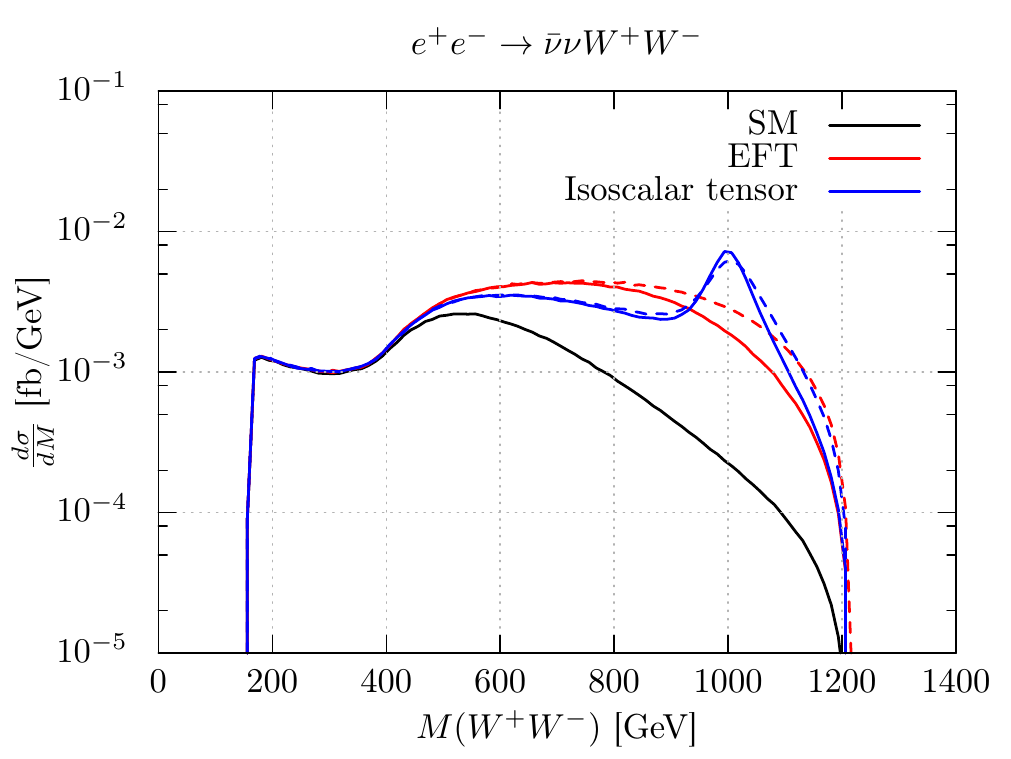}
   \end{subfigure}\hfill%
   \begin{subfigure}[t]{0.5\textwidth}
      \includegraphics[width=\textwidth]{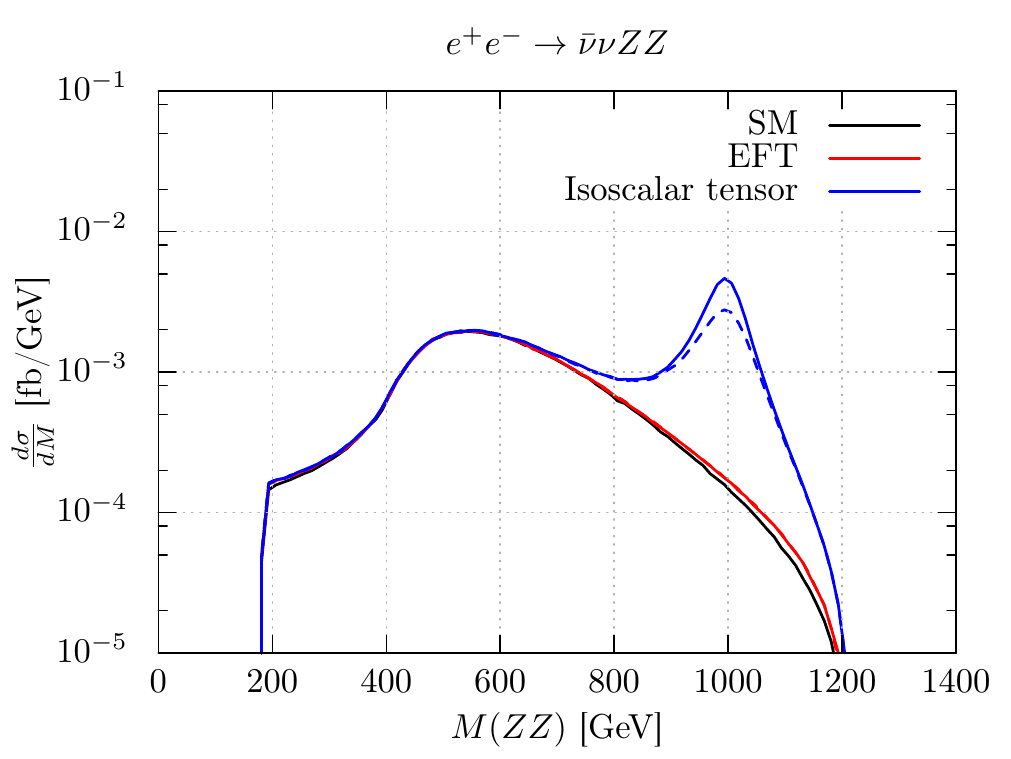}
   \end{subfigure}\\[5pt]%
   \centering
   \begin{subfigure}[t]{0.5\textwidth}
      \includegraphics[width=\textwidth]{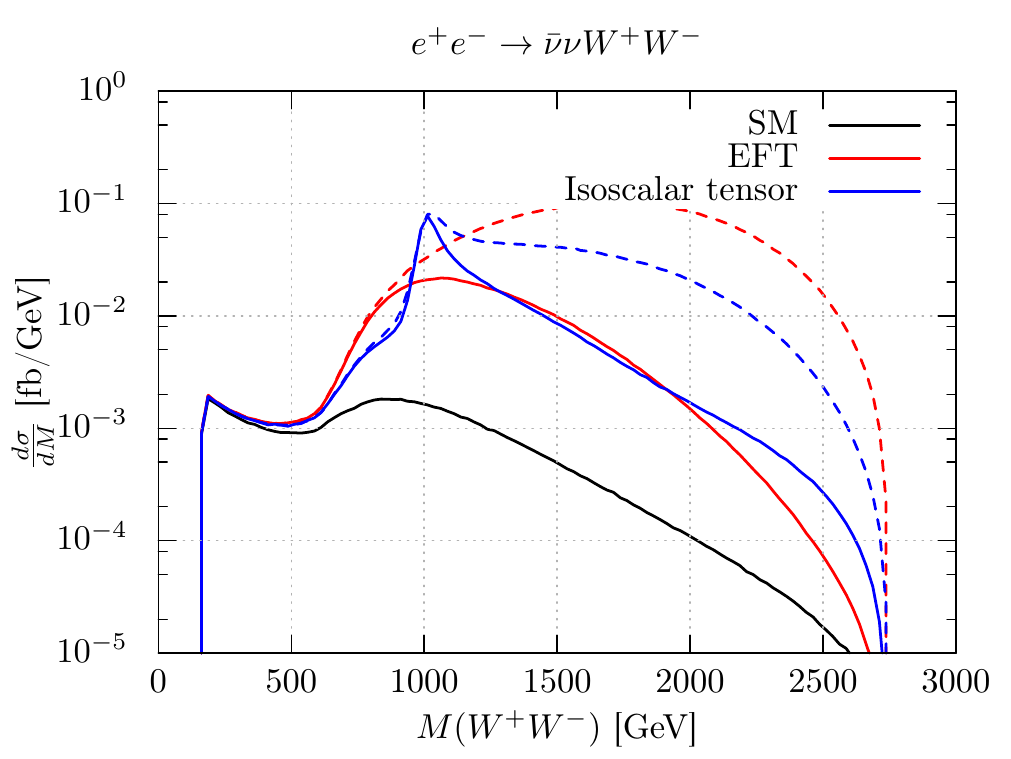}
   \end{subfigure}\hfill%
   \begin{subfigure}[t]{0.5\textwidth}
      \includegraphics[width=\textwidth]{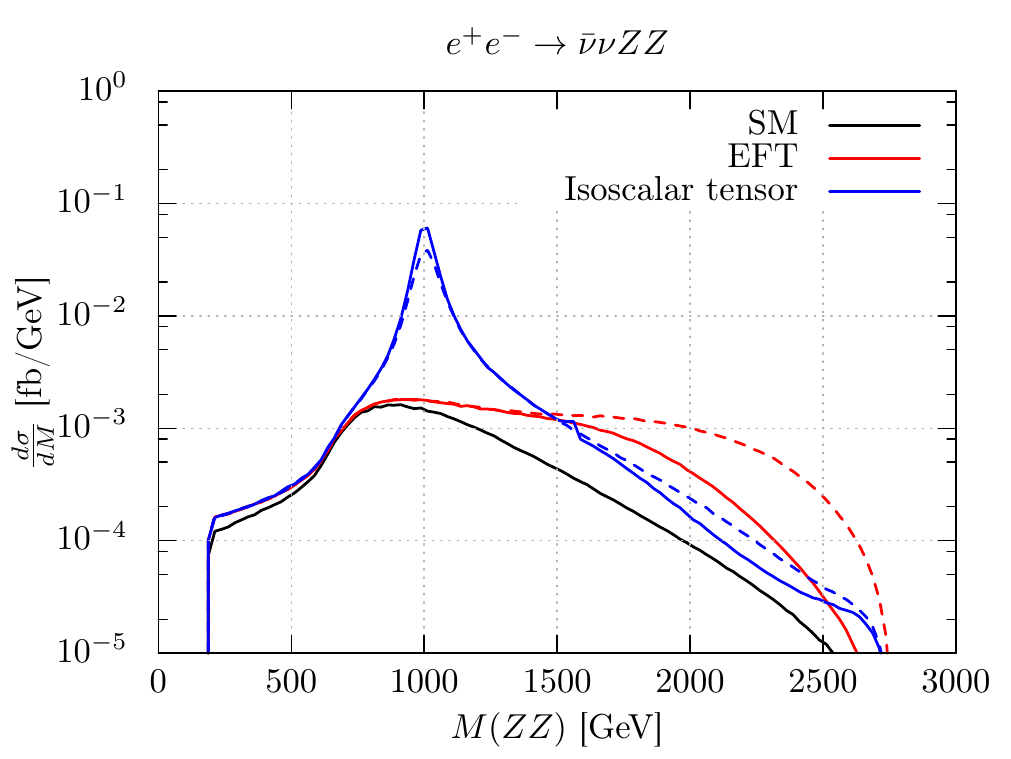}
   \end{subfigure}
   \caption{Differential cross sections including an isoscalar
            tensor resonance (${m_f=1000\;\GeV}$, $F_f=17.4\;\TeV^{-1}$,
            $\Gamma_f=100\;\GeV$) depending on the invariant mass of
            the vector boson pair at center-of-mass energies of
            $\sqrt{s}=1.4\;\TeV$ (upper plots) and $\sqrt{s}=3\;\TeV$
            (lower plots). Plots on the left show the process
            $e^+e^-\to \bar{\nu}\nu W^+W^-$, plots on the right
            $e^+e^-\to \bar{\nu}\nu ZZ$. Blue line: isoscalar tensor
            resonance, red line: matched EFT results
            ($F_{S,0}=150.8\;\TeV^{-4}$, $F_{S,1}=-50.3\;\TeV^{-4}$).
            Solid line: unitarized results, dashed line: naive results.}
   \label{i:isoscalar_tensor_1}
\end{figure}

In Fig~\ref{i:isoscalar_tensor_1}, we consider a isoscalar tensor
resonance with $M_f=1\;\TeV$ and $\Gamma_f=100\;\GeV$, which produces a
rather narrow peak in all distributions.  With a collider energy of
$3\;\TeV$, we observe the necessity for unitarization beyond the mass
peak (in the $WW$ final state), caused by the dimensionality of the
effective tensor-scalar interaction.  As well as in all other cases,
if we had a UV-complete model at hand, we would expect any variation
of the resonance-model prediction (blue) in this range: further
resonances, a featureless continuum, or suppression that accommodates
the emergence of further inelastic channels.  However, neither of these
scenarios could produce a unitarity-violating result like the naive
blue-dashed line, so the unitarized model prediction serves as a
conservative estimate of the asymptotic shape.

\begin{figure}[p]
   \begin{subfigure}[t]{0.5\textwidth}
      \includegraphics[width=\textwidth]{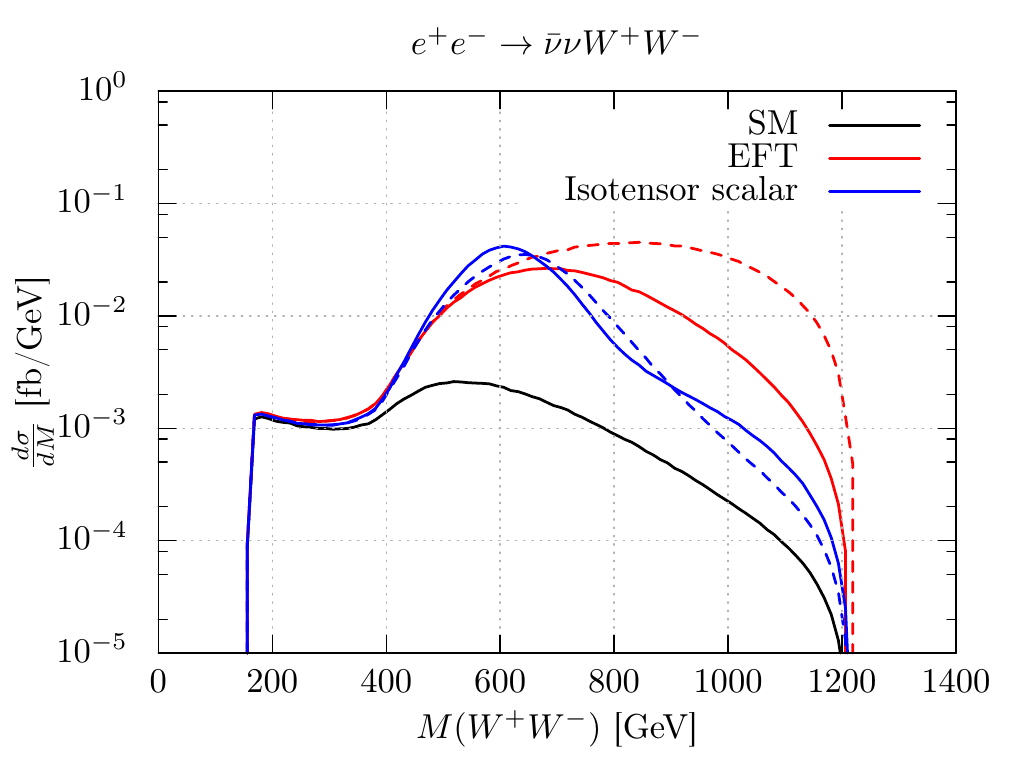}
   \end{subfigure}\hfill%
   \begin{subfigure}[t]{0.5\textwidth}
      \includegraphics[width=\textwidth]{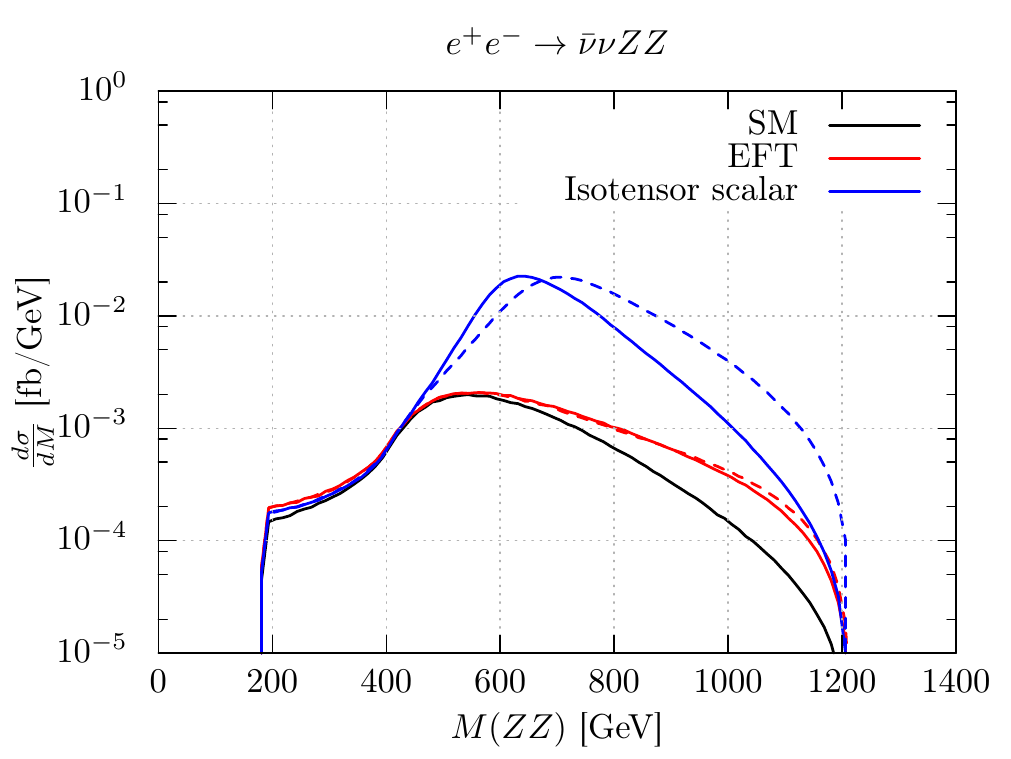}
   \end{subfigure}\\[5pt]%
   \centering
   \begin{subfigure}[t]{0.5\textwidth}
      \includegraphics[width=\textwidth]{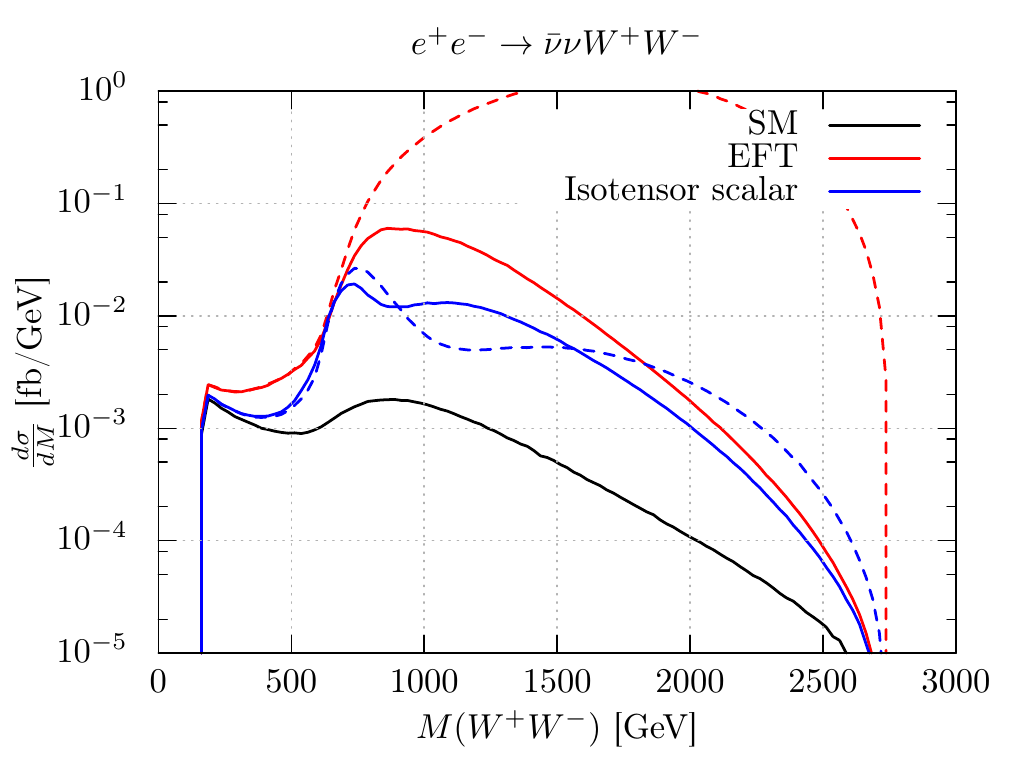}
   \end{subfigure}\hfill%
   \begin{subfigure}[t]{0.5\textwidth}
      \includegraphics[width=\textwidth]{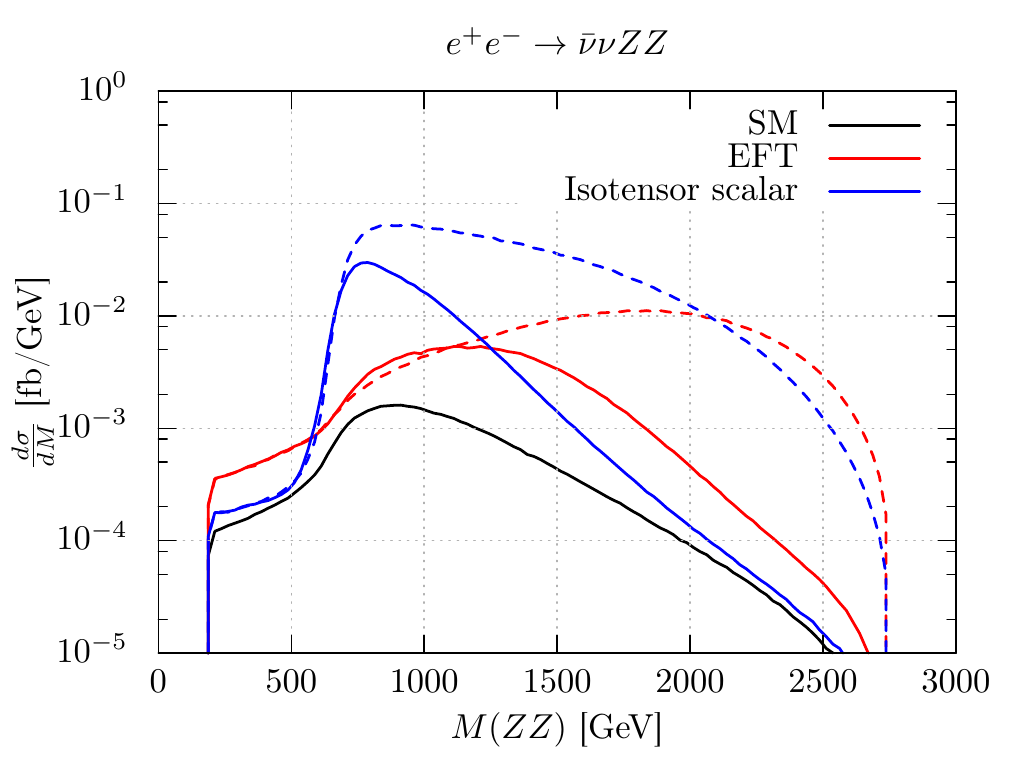}
   \end{subfigure}
   \caption{Differential cross sections including a low lying isotensor
            scalar resonance (${m_\phi=650\;\GeV}$, $F_\phi=19.5\;\TeV^{-1}$,
            $\Gamma_\phi=260\;\GeV$) depending on the invariant mass of
            the vector boson pair at center-of-mass energies of
            $\sqrt{s}=1.4\;\TeV$ (upper plots) and $\sqrt{s}=3\;\TeV$
            (lower plots). Plots on the left show the process
            $e^+e^-\to \bar{\nu}\nu W^+W^-$, plots on the right
            $e^+e^-\to \bar{\nu}\nu ZZ$. Blue line: isotensor scalar
            resonance, red line: matched EFT results
            ($F_{S,0}=450.5\;\TeV^{-4}$,$F_{S,1}=-112.6\;\TeV^{-4}$).
            Solid line: unitarized results, dashed line: naive results.}
   \label{i:isotensor_scalar_2}
\end{figure}

The final parameter set in Fig.~\ref{i:isotensor_scalar_2}, an
isotensor-scalar multiplet $\phi$, illustrates a possible strongly
interacting multi-Higgs scenario.  The broad resonance, actually a
combination of resonance exchange in all isospin channels, is
indistinguishable from an arbitrary continuum.  It is remarkable that
the EFT approximation follows the shape of the resonance model in the
$WW$ final state but fails completely for the $ZZ$ final state.  As
before, the naive extrapolations (dashed) overshoot the unitarized
models (solid) by a large amount.

\section{Conclusions}
We have performed a new study of the capability of a high-energy
lepton collider (such as CLIC or an upgraded ILC) to measure
quasi-elastic vector-boson scattering, as a dedicated probe of the
Higgs sector.  For realistic luminosity-energy combinations, we cannot
restrict the investigation to a pure effective field theory (EFT), but have
to take into account strong interactions or resonant behavior.
Specifically, we have considered a straightforwardly extrapolated EFT
Lagrangian and related this to alternative models where the anomalous
effects develop into resonances within the kinematical range.

Our phenomenological scenarios all reduce to the Standard Model with
leading higher-dimensional corrections at the low-energy threshold for
vector boson scattering processes, but run into unitarity saturation
at high energy or on a resonance peak.  The T-matrix unitarization
framework lets us handle this issue by a universal algorithm and
simultaneously allows us to embed the models in a full off-shell
calculation and Monte-Carlo simulation.  For the purpose of working
with fully exclusive event samples, all models have been
implemented in the public version of the WHIZARD Monte-Carlo event
generator.

The results yield an overview over the experimental reach of
the CLIC and ILC colliders in the various scenarios.  Using optimized
cuts for longitudinal vector boson scattering, we generically obtain a
sensitivity on the extrapolated EFT parameters $F_{S,0}$ and $F_{S,1}$
of $O(1/\Lambda^4)$, where $\Lambda$ indicates the location of the
maximum of the unitarized cross section, in terms of the
vector-boson pair invariant mass as the relevant energy scale.
Regarding resonance models, we have restricted ourselves to a few
exemplary parameter points in this exploratory study, which should be
representative of the phenomenology that can be expected.  For precise and systematic parameter determinations, the
specific models should be subject to an experimental fitting procedure and
analysis which takes into account complete event information and
weighs the distributions according to their sensitivity to the
specific model in question.

The expected sensitivity of the CLIC collider with $3\,\TeV$, as expressed for the
parameters of the extrapolated EFT as the simplest model, improves over the current LHC
limits by two orders of magnitude.  Clearly, the CLIC expectation has
to be compared to the ultimate precision achievable by the LHC
experiments.  A detailed comparison of sensitivities would require
applying the available
analysis techniques to a full-simulation prediction for either
collider, which is beyond the scope of the present paper.  However,
the $e^+e^-$ environment allows for the detection of
well-defined final states, hadronic decays, and a direct measurement
of the most relevant distribution -- the vector-boson pair invariant
mass -- and thus remains the preferred setup for a comprehensive study of VBS in the
$\TeV$ energy range.

\section*{Acknowledgments}

We would like to thank Lucie Linssen, Philipp Roloff and Marcel Vos for
enlightening discussions on the CLIC project. MS acknowledges the
support of BMBF Verbundforschung (HEP Theory).

\clearpage

\appendix

\section{Dimension-8 Operators that Affect VBS}

The following list of dimension-8 operators includes all leading interactions
that modify the SM form of VBS interactions.  In the current paper, we
restrict the investigation to the first set of operators which
describes genuine Goldstone-Higgs self-interactions, which for
completeness we repeat here from~\ref{eq:dim8_operator_S}:
\begin{subequations}
  \begin{alignat}{4}
    \LL_{S,0}&=
    & &F_{S,0}\ &&
    \tr{ \left ( \vD_\mu \vH \right )^\dagger \vD_\nu \vH}
    \tr{ \left ( \vD^\mu \vH \right )^\dagger \vD^\nu \vH},
    \\
    \LL_{S,1}&=
    & &F_{S,1}\ &&
    \tr{ \left ( \vD_\mu \vH \right )^\dagger \vD^\mu \vH}
    \tr{ \left ( \vD_\nu \vH \right )^\dagger \vD^\nu \vH}.
  \end{alignat}
\end{subequations}
Exchanging two covariant derivatives with field strength tensors,
further possibilities arise to construct linearly independent
operators
\begin{subequations}
  \label{eq:dim8_operator_M}
  \begin{alignat}{3}
    \LL_{M,0}&=&-g^2&F_{M,0}& &\tr{ \left( \vD_\mu \vH \right)^\dagger
      \left( \vD^\mu \vH \right)}
    \tr{\vW_{\nu\rho}
      \vW^{\nu\rho}} , \\
    \LL_{M,1}&=&-g^2&F_{M,1}& &\tr{ \left( \vD_\mu \vH \right)^\dagger
      \left( \vD^\rho \vH \right)}
    \tr{\vW_{\nu\rho}
      \vW^{\nu\mu}} , \\
    \LL_{M,2}&=&-{g^\prime}^2&F_{M,2}& &\tr{ \left( \vD_\mu \vH \right)^\dagger
      \left( \vD^\mu \vH \right)}
    \tr{\vB_{\nu\rho}
      \vB^{\nu\rho}} , \\
    \LL_{M,3}&=&-{g^\prime}^2&F_{M,3}& &\tr{ \left( \vD_\mu \vH \right)^\dagger
      \left( \vD^\rho \vH \right)}
    \tr{\vB_{\nu\rho}
      \vB^{\nu\mu}} , \\
    \LL_{M,4}&=&-gg^\prime&F_{M,4}& &\tr{ \left( \vD_\mu \vH \right)^\dagger
      \vW_{\nu\rho}
      \left( \vD^\mu \vH \right)
      \vB^{\nu\rho}} , \\
    \LL_{M,5}&=&-gg^\prime&F_{M,5}& &\tr{ \left( \vD_\mu \vH \right)^\dagger
      \vW_{\nu\rho}
      \left( \vD^\rho \vH \right)
      \vB^{\nu\mu}} , \\
    \LL_{M,7}&=&-g^2&F_{M,7}& &\tr{ \left( \vD_\mu \vH \right)^\dagger
      \vW_{\nu\rho}
      \vW^{\nu\mu}
      \left( \vD^\rho \vH \right)}.
  \end{alignat}
\end{subequations}
Here, we kept the numbering analogue to the linear Higgs doublet
representation in~\cite{Baak:2013fwa}  for future comparisons, where
some linear dependent operators of~\cite{Eboli:2006wa} are already
omitted.

Operators affecting only the gauge bosons consist of four
electroweak field strength tensors
\begin{subequations}
  \label{eq:dim8_operator_T}
  \begin{alignat}{3}
    \LL_{T,0}&=&g^4&F_{T,0}& &\tr{\vW_{\mu\nu}\vW^{\mu\nu}}
    \tr{\vW_{\alpha\beta}\vW^{\alpha\beta}} , \\
    \LL_{T,1}&=&g^4&F_{T,1}& &\tr{\vW_{\alpha\nu}\vW^{\mu\beta}}
    \tr{\vW_{\mu\beta}\vW^{\alpha\nu}} , \\
    \LL_{T,2}&=&g^4&F_{T,2}& &\tr{\vW_{\alpha\mu}\vW^{\mu\beta}}
    \tr{\vW_{\beta\nu}\vW^{\nu\alpha}} , \\
    \LL_{T,5}&=&g^4&F_{T,5}& &\tr{\vW_{\mu\nu}\vW^{\mu\nu}}
    \tr{\vB_{\alpha\beta}\vB^{\alpha\beta}} , \\
    \LL_{T,6}&=&g^4&F_{T,6}& &\tr{\vW_{\alpha\nu}\vW^{\mu\beta}}
    \tr{\vB_{\mu\beta}\vB^{\alpha\nu}} , \\
    \LL_{T,7}&=&g^4&F_{T,7}& &\tr{\vW_{\alpha\mu}\vW^{\mu\beta}}
    \tr{\vB_{\beta\nu}\vB^{\nu\alpha}} , \\
    \LL_{T,8}&=&{g^\prime}^4&F_{T,8}& &\tr{\vB_{\mu\nu}\vB^{\mu\nu}}
    \tr{\vB_{\alpha\beta}\vB^{\alpha\beta}} , \\
    \LL_{T,9}&=&{g^\prime}^4&F_{T,9}& &\tr{\vB_{\alpha\mu}\vB^{\mu\beta}}
    \tr{\vB_{\beta\nu}\vB^{\nu\alpha}}.
  \end{alignat}
\end{subequations}

\clearpage
\bibliographystyle{unsrt}

\end{document}